\documentclass[twocolumn,showpacs,preprintnumbers,amsmath,amssymb]{revtex4}

\usepackage{graphicx}
\usepackage{dcolumn}
\usepackage{bm}

\newcommand{\fex}{f^{\rm ex}} 
\newcommand{\pa}{^{(\alpha)}} 
\newcommand{\eq}[1]{~(\ref{#1})} 
\newcommand{\fmom}{f_{\rm mom}}
\newcommand{\be}{\begin{equation}}
\newcommand{\ee}{\end{equation}}

\newcommand{\rh}{\rho}  
\newcommand{\sig}{\sigma} 
\newcommand{\rhsig}{\rh(\sig)} 
\newcommand{\rhzsig}{\rh^{(0)}(\sig)} 
\newcommand{\nsig}{n(\sig)} 
\newcommand{\pol}{\delta} 
\newcommand{\polt}{\pol_{\rm t}}
\newcommand{\prior}{R(\sig)}
\newcommand{\unil}{\sig_0} 
\newcommand{\univ}{v_0} 
\newcommand{\til}[1]{\tilde{#1}} 

\newcommand{\muexsig}{\mu^{\rm ex}(\sig)} 
\newcommand{\musig}{\mu(\sig)} 
\newcommand{\muex}{\mu^{\rm ex}} 
\newcommand{\intsig}{\int\!d\sig\,}

\newcommand{\muze}{\mu_0}  
\newcommand{\muon}{\mu_1} 
\newcommand{\mutw}{\mu_2} 
\newcommand{\muth}{\mu_3} 

\newcommand{\wi}{w_i(\sig)} 

\newcommand{\mi}{\rh_i} 
\newcommand{\mze}{\rh_0}
\newcommand{\mon}{\rh_1}
\newcommand{\mtw}{\rh_2}
\newcommand{\mth}{\rh_3}

\newcommand{\cwi}{[(\sig-1)/\pol_0]^i} 

\newcommand{\ctr}{^{\rm c}}

\begin{document}


\title{Fractionation effects in phase equilibria of polydisperse hard sphere colloids}
       
\author{Moreno Fasolo}
 \email{moreno.fasolo@kcl.ac.uk}
\author{Peter Sollich}
 \email{peter.sollich@kcl.ac.uk}
\affiliation{
Department of Mathematics, King's College London, London WC2R 2LS, U.K.
}

\date{\today}

\begin{abstract}
The equilibrium phase behaviour of hard spheres with size
polydispersity is studied theoretically. We solve numerically the
exact phase equilibrium equations that result from accurate free
energy expressions for the fluid and solid phases, while accounting
fully for size fractionation between coexisting phases. Fluids up to
the largest polydispersities that we can study (around 14\%) can phase
separate by splitting off a solid with a much narrower size
distribution.  This shows that experimentally observed terminal
polydispersities above which phase separation no longer occurs must be
due to non-equilibrium effects. We find no evidence of re-entrant
melting; instead, sufficiently compressed solids phase separate into
two or more solid phases. Under appropriate conditions, coexistence of
multiple solids with a fluid phase is also predicted. The solids have
smaller polydispersities than the parent phase as expected, while the
reverse is true for the fluid phase, which contains predominantly
smaller particles but also residual amounts of the larger ones.
The properties of the coexisting phases are studied in detail; mean
diameter, polydispersity and volume fraction of the phases all reveal
marked fractionation. We also propose a method for constructing
quantities that optimally distinguish between the coexisting phases,
using Principal Component Analysis in the space of density
distributions.
We conclude by comparing our predictions to perturbative
theories for near-monodisperse systems and to Monte Carlo simulations
at imposed chemical potential distribution, and find excellent
agreement.
\end{abstract}



\maketitle


\section{\label{sec:3hsintroduction}Introduction}

\subsection{\label{sec:3hsthehardspheremodel}The hard sphere model}

Hard spheres are particles that do not interact except via an infinite
repulsion on contact.
In a hard sphere system there is no contribution to the internal
energy, $U$, from interparticle forces since $U$ is zero for all the
allowed configurations. Minimising the free energy, $F=U-TS$, is thus
equivalent to maximising the entropy, $S$: the structure and phase
behaviour of hard spheres is determined solely by entropy. Temperature
$T$ only features as a trivial factor setting the energy scale.

The hard sphere model was originally introduced as a mathematically
simple model of atomic liquids (see e.g.~\cite{HanMcD86}), but has
since also been recognised as a useful basic model for complex
fluids~\cite{RusSavSch89} such as spherical \emph{colloids}.
Colloidal particles coated with a thin polymeric layer so that strong
steric repulsions dominate the attractive dispersion forces between
the colloidal cores behave in many ways as hard spheres. Indeed,
crystallisation can be observed at densities similar to those
predicted by computer simulation for hard spheres, with a single-phase
fluid below volume fractions $\phi\approx 0.494$, fluid-solid
coexistence at up to $\phi\approx 0.545$, and a single-phase solid at
higher volume fractions~\cite{PusVan86,PauAck90}. Measurements of the
osmotic pressure and compressibility similarly show very good
agreement with predicted hard sphere properties~\cite{Goetze91}.

There is, however, one important and unavoidable difference between
colloids and the classical hard sphere model: whereas the spheres in
the classical model are identically sized, colloidal particles have an
inevitable spread of diameters.
The magnitude of this spread is conveniently characterised by the
parameter $\pol$, which is often also referred to as polydispersity
and measures the standard deviation of the diameter distribution
normalised by its mean:
\begin{equation}
\pol = \frac{\left(\overline{\sig^{2}} - 
\overline{\sig}^{2} \right)^{\frac{1}{2}}}{\overline{\sig}} \; .
\label{eq:pol}
\end{equation}
Here the averages $\overline{\sig}$ and $\overline{\sig^2}$ are
defined via
\begin{equation}
\mi \equiv \rh\overline{\sig^{i}} = \int d\sig \rhsig
\sig^{i} \; , 
\end{equation}
with $\rhsig$ the {\em density distribution} of the system. The latter
is defined so that the number density of particles with diameter
between $\sig$ and $\sig+d\sig$ is given by $\rhsig\;d\sig$. The total
density is then $\rh = \int d\sig \rhsig$, and $n(\sig)=\rhsig/\rh$
is the normalised diameter distribution. The $\mi$ are the moments
of the density distribution, with $\mze\equiv\rh$.
The presence of polydispersity in the system brings in a new
parameter that allows us to distinguish between size distributions of
different widths; the shape of the diameter distribution is of course
also relevant. Compared to the monodisperse case, polydispersity
causes several qualitatively new phenomena which have
received much interest in recent years.

\subsection{\label{sec:3hsnewphenomenaarisingfrompolydispersity}New
phenomena arising from polydispersity}

The effect of polydispersity on the phase behaviour of hard spheres
has been investigated by experiments~\cite{PusVan86,Pusey91}, computer
simulations~\cite{DicPar85,BolKof96,BolKof96b,PhaRusZhuCha98,KofBol99},
density functional theories~\cite{BarHan86,McrHay88}, and simplified
analytical
theories~\cite{Pusey87,PhaRusZhuCha98,Bartlett97,Bartlett98,%
Sear98,BarWar99,XuBau03,Sollich02}. We will now outline the main
findings and introduce the relevant terminology.

First, it is intuitively clear~\cite{Pusey87} that significant
diameter polydispersity should {\it destabilize the crystal phase},
because it is difficult to accommodate a range of diameters in a
lattice structure. Experiments have indeed shown that crystallisation
is suppressed above a {\em terminal polydispersity} of $\polt \approx
0.12$~\cite{PusVan86,Pusey91}.
Since then much theoretical work has focused on estimating
$\polt$. Dickinson {\em et al}~\cite{DicPar85}, for example,
extrapolated the decrease of the volume change on melting with
polydispersity to zero, obtaining an estimate of $\polt\approx
0.12$. Pusey~\cite{Pusey87} used a simple Lindemann-type criterion to
estimate that the larger spheres in a polydisperse system would
disrupt the crystal structure above $\polt\approx 0.06\ldots0.12$. McRae
and Haymet~\cite{McrHay88} used density functional theory (DFT) and
found that there was no crystallisation above $\polt\approx
0.05$. Barrat and Hansen~\cite{BarHan86} also employed DFT, estimating
the free energy difference between fluid and solid.
Taken together, this body of theoretical work suggests that the
terminal polydispersity arises from a progressive narrowing of the
fluid-solid coexistence region with increasing $\pol$, with phase
boundaries meeting at $\polt$~\cite{McrHay88,Bartlett97} in a point
that has been identified as one of equal concentration~\cite{BarWar99}
(rather than a critical point). 
\begin{figure}
\begin{center}
\includegraphics[width=5cm]{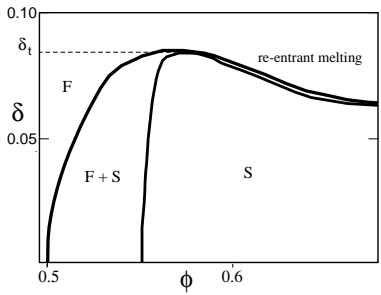}%
\caption{Sketch of the fluid (F) and solid (S) phase boundaries for
polydisperse hard spheres, following~\cite{BarWar99}. The boundaries
are plotted as polydispersity $\pol$ versus volume fraction $\phi$.
The fluid boundary approaches the solid one until they
meet at a \emph{terminal polydispersity}, $\polt$. For $\pol$ just
below $\polt$, this scenario suggests {\em re-entrant melting}:
compressing the crystal to sufficiently high volume fraction should
transform it back into a fluid. 
\label{fig:pha_dia_ske}
}
\end{center}
\end{figure}

%
Bartlett and Warren~\cite{BarWar99} also found {\em re-entrant
melting} on the high-density side of this point: for $\pol$ just below
$\polt$, they predicted that compressing a crystal could transform it
back into a fluid. Fig.~\ref{fig:pha_dia_ske} shows a
sketch of this scenario.

Physically, the existence of re-entrant melting would suggest that,
while in the monodisperse case the solid has the lower free energy at
all volume fractions above $\phi \approx 55$\%, the fluid can become
preferred again at large $\phi$ if the polydispersity is sufficiently
large. This result is compatible with the intuition that
polydispersity {\em reduces} the maximum packing fraction in a crystal
(since a range of diameters need to be accommodated on uniformly
spaced lattice sites), while it {\em increases} the maximum packing
fraction in the fluid, where smaller spheres should be able to fill
``holes'' between larger particles more easily.

\begin{figure}
\begin{center}
	\includegraphics[width=8.5cm]{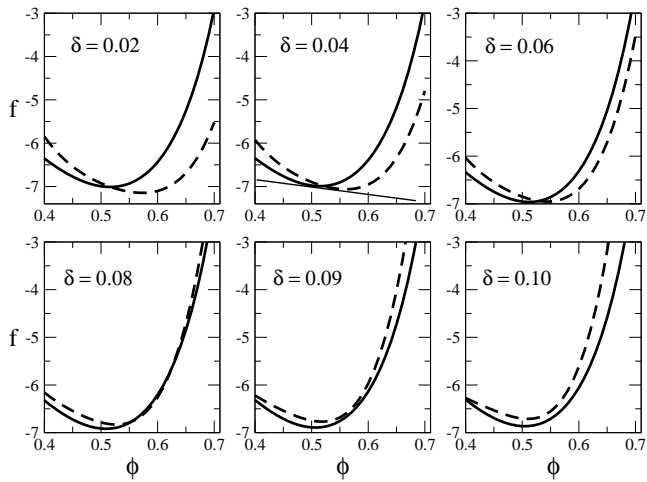}%
	\caption{Free energy $f$ versus $\phi$ within the ``no
	fractionation'' approximation used by Bartlett and
	Warren~\protect\cite{BarWar99}. Phase separation occurs where
	double tangents between the fluid (thick line) and solid
	(dashed line) branches of the free energy can be drawn; the
	plot at $\pol=0.04$ shows an example (thin line). At
	$\pol=0.08$ re-entrant melting can be observed: two double
	tangents can now be drawn. For larger $\pol$, phase separation
	is no longer predicted. (A linear term $-19\phi$ has been
	added to all free energies to make the plots more readable.)
\label{fig:barwartpd}}
\end{center}
\end{figure}

This intuition can be made more quantitative by comparing the fluid
and solid free energies, following~\cite{Bartlett00}. The basic
analysis by Bartlett and Warren~\cite{BarWar99} ignores fractionation,
i.e.\ the fact that coexisting phases need not have identical diameter
distributions as long as they combine to give the correct overall or
``parent'' distribution $\rhzsig$. The normalised diameter
distribution is thus fixed and equal in all phases. In the moment free
energy (MFE) method described below this corresponds to retaining only
the overall density $\mze$. Phase equilibria can then be found by the
usual double-tangent construction~\cite{Sollich02} from a plot of the
(moment) free energy density $f$ versus $\mze$.
We display such free energy plots in Fig.~\ref{fig:barwartpd}, showing
along the $x$-axis the volume fraction $\phi$ rather than $\mze$; the
two are proportional for fixed diameter distribution. (The free
energies are those also used for our detailed calculations below. The
diameter distribution was of a Schultz form, but other size
distributions are expected to give similar results.)
For polydispersities up to $\pol=0.08$ a tangent plane between the
solid and fluid phases always exists, giving the conventional
fluid-solid coexistence.  At $\pol=0.08$, re-entrant melting has
appeared: a second double tangent is possible because at large volume
fractions the solid free energy is now {\em higher} than that of the
fluid. As $\pol$ increases, the solid free energy continues to
increase relative to the fluid and eventually lies above the latter
for all $\phi$ (see $\pol=0.09$ and 0.1 in
Fig.~\ref{fig:barwartpd}). The point where this first happens gives
the terminal polydispersity $\polt$; in our example $\polt \approx
0.083$. As $\pol$ 
approaches $\polt$ from below, the widths of both the ordinary and the
re-entrant fluid-solid coexistence regions shrink to zero and merge
into the point of equal concentration, consistent with the phase
diagram shown in Fig.~\ref{fig:pha_dia_ske}.

%
%

As mentioned above, this picture ignores the possibility
of fractionation. Bartlett and Warren~\cite{BarWar99} investigated
fractionation effects approximately, by using a MFE with {\em two}
density variables included, $\mze$ and $\mon = \rh\bar{\sig}$%
. They concluded that the phase diagram topology remained
qualitatively unchanged; quantitatively, the point of equal
concentration was shifted to higher density and lower
polydispersity. It has to be born in mind, however, that while the
approach of~\cite{BarWar99} allowed coexisting phases to have
different mean diameters, it implicitly still constrained them to have
the same $\pol$. (This is because, within the MFE method applied to
the Schultz prior $R(\sig)\propto \sig^z e^{-(z+1)\sig}$
of~\cite{BarWar99}, the density distributions
$\rhsig=R(\sig)e^{\lambda_0+\lambda_1\sig} \propto
\sig^z e^{[\lambda_1-(z+1)]\sig}$ in all phases are also Schultz, with
common $z$ and therefore common $\pol=(1+z)^{-1/2}$.)
On the other hand, numerical simulations that allow for fractionation
show that a solid with a narrow size distribution can coexist with an
essentially arbitrarily polydisperse fluid~\cite{BolKof96,BolKof96b,KofBol99}.
This suggests that the prediction of re-entrant melting should be
re-examined theoretically, allowing for such fractionation
effects. Conceptually, it
also implies that the concept of a terminal polydispersity is likely
to be useful only for the solid but not for the fluid, and we will see
this confirmed below.

\begin{figure}
\begin{center}
	\includegraphics[width=7cm]{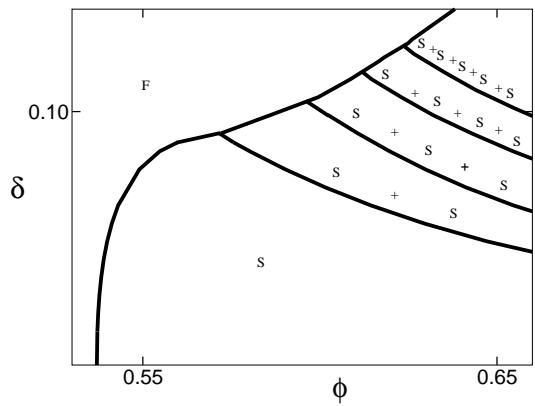}%
	\caption{Sketch of fluid (F) and multiple solid (S) phase
	coexistences in polydisperse hard spheres,
	following~\protect\cite{Bartlett98}. Approximate phase
	boundaries are plotted as polydispersity $\pol$ versus volume
	fraction $\phi$.
        For sufficiently large $\pol$ and $\phi$, coexistence of
        several solids is predicted; see text for discussion.
\label{fig:solids}
}
\end{center}
\end{figure}

Fractionation has also been predicted to lead to {\em solid-solid
coexistence}~\cite{Bartlett98,Sear98,Bartlett00}, where a broad diameter
distribution is split into a number of narrower solid fractions. This
occurs because the loss of entropy of mixing is outweighed by the
better packing, and therefore higher entropy, of crystals with narrow
size distribution; accordingly, as the overall polydispersity of the
system grows, the number of coexisting solids is predicted to
increase. 
Fig.~\ref{fig:solids} sketches this effect, following the treatment
of~\cite{Bartlett98}. There is no coexistence region between fluid and
solid, due to a simplification in the analysis of~\cite{Bartlett98}:
rather than solving the phase equilibrium conditions, only the free
energies were equated between the fluid and the (one or several) solid
phases. The resulting lines in the phase diagram generally lie inside
the actual phase separation region, but give a rough guide to the
phase transitions that can occur. The parent diameter distribution
considered had a ``top hat'' form (uniform between given minimum and
maximum diameters), and for multiple solids fractionation was assumed
to be ``hard'', with the parent distribution split into 
non-overlapping top hat distributions with identical
polydispersities.

%
Previous work as described above leaves open a number of questions.
The rather drastic, and differing, approximations for size
fractionation used in previous studies of re-entrant melting and
solid-solid coexistence~\cite{BarWar99,Bartlett98,Sear98,Bartlett00},
as described above, leave the relative importance of these two
phenomena unclear.
Theoretical calculations that account fully for fractionation remain
restricted to highly simplified van der Waals free
energies~\cite{XuBau03}. Numerical simulations can in principle also
capture arbitrary fractionation behaviour, but have been carried out
at constant chemical potential
distribution~\cite{BolKof96,BolKof96b,KofBol99}.  As explained in more
detail in Section~\ref{sec:3hsComparisonwithBolhuisandKofkeresults},
the system's overall particle size distribution can then change
dramatically across the phase diagram. This is in contrast to the
experimental situation and so limits the applicability of the results.

Our main aim in this study is, therefore, to calculate the equilibrium
phase behaviour of polydisperse hard spheres on the basis of accurate
free energy expressions, taking full account of fractionation and
going beyond previous work on fluid-solid and solid-solid coexistence.
The experimentally observed behaviour of hard sphere colloids will of
course also depend on {\it non-equilibrium effects}, e.g.\ the
presence of a kinetic glass transition~\cite{PusVan87}, anomalously
large nucleation barriers~\cite{AueFre01} or the growth kinetics of
polydisperse crystals~\cite{EvaHol01}. Nevertheless, the equilibrium
phase behaviour needs to be understood as a baseline from which
non-equilibrium effects can be properly attributed. Also, more of the
equilibrium behaviour may be observable under microgravity conditions,
where the glass transition is shifted to higher densities or even
absent~\cite{ZhuLiRogMeyOttRusCha97}.

We begin in Section~\ref{sec:3hsfreeenergychoice} by defining the free
energies we use to describe the fluid and solid phases of polydisperse
hard spheres. 
Section~\ref{sec:MFE} reviews the moment free energy method and its
numerical implementation for solving the phase equilibrium conditions.
In Section~\ref{sec:phasediag} we then describe the basic features of
the phase behaviour that we find; a short account of these results has
appeared
in~\cite{FasSol03}. Section~\ref{sec:3hsFractionationbehaviour}
describes in detail the fractionation effects that we predict, and
introduces a new method for constructing optimal visualisations of
polydisperse phase behaviour. Section~\ref{sec:comparison}, finally,
compares our results to perturbative theories for the
near-monodisperse limit and to Monte Carlo simulations at constant
chemical potential differences. The agreement is very good, thus
validating our approach. We conclude in
Section~\ref{sec:conclusion} with a summary and outlook towards future
work.

\section{\label{sec:3hsfreeenergychoice}Free energies}

Our starting point is the decomposition of the free energy of a
polydisperse system into an ideal and an excess part,
\begin{equation}
f = \int d\sig \rhsig \left[\ln\rhsig-1\right] +
\fex(\{\mi\})
\label{eq:fre_ene}
\end{equation}
The excess part $\fex$ can, in principle, depend on all details of
$\rhsig$ and therefore on all of its moments $\mi$, but we will be
concerned with {\em truncatable} free energies~\cite{SolWarCat01}. For
these, the dependence is only through a finite number of moments, for
us specifically $\mze,\ldots, \mth$.

Strictly speaking, equation (\ref{eq:fre_ene}) gives the free energy
{\em density}; we will continue to refer to this as the free energy
for short. Also, all quantities in (\ref{eq:fre_ene}) are
dimensionless: we assume that sphere diameters are measured in units
of some reference value $\unil$, that all densities are
made dimensionless by multiplying by the volume
$\univ=\pi\unil^3/6$
of a reference sphere, and that all energies are measured in units of
$T=1/\beta$. Boltzman's constant $k_{\rm B}$ is set to 1
throughout. Free energy and pressure are then in units of $T/\univ$,
for example. Table~\ref{tab:dimquantity} summarises the relations
between important dimensionless and dimensional quantities%
. Conveniently, with our choice of units $\mth\equiv\phi$ is simply
the volume fraction of spheres.
%

\begin{table}
\begin{center}
\begin{tabular}{|ccc|}
\hline
Dimensionless & & Dimensional \\
\hline
$\sig$                   & = & $\til{\sig}/\sig_0$\\
$f$                      & = & $\beta \univ \til{f}$\\
$P$                      & = & $\beta\univ \til{P}$\\
$\rhsig$                 & = & $\univ \sig _0 \til{\rh}(\til{\sig})$\\
$\mi$                    & = & $({\univ}/{\unil^i})\til{\rh}_i$\\
$\muex(\sig)$            & = & $\beta\til{\mu}^{\rm ex}(\til{\sig})$\\
$\mu^{\rm ex}_i$         & = & $\unil^i\til{\mu}_i^{\rm ex}$\\
\hline
\end{tabular}
\end{center}
\caption{Relations between dimensional and dimensionless
quantities. All dimensional quantities except for the units $\beta$, $\unil$
and $\univ$ themselves are denoted by tildes ``$\sim$''.
\label{tab:dimquantity}
}
\end{table}


%
For the {\it fluid} phase of polydisperse hard spheres, the most
accurate free energy approximation available is the generalisation by
Salacuse and Stell~\cite{SalSte82} of the equation of state due
to Boublik, Mansoori, Carnahan, Starling and
Leland (BMCSL)~\cite{Boublik70,ManCarStaLel71}; for the monodisperse case this
reproduces the Carnahan-Starling equation of state~\cite{CarSta69}.
In our dimensionless quantities, the BMCSL expression for the excess
free energy takes the form
\begin{equation}
\fex = \left(\frac{\mtw^3}{\mth^2} - \mze\right)\ln(1-\mth) +
\frac{3 \mon \mtw}{1-\mth} + \frac{\mtw^3}{\mth(1-\mth)^2}
\label{eq:BMCSL}
\end{equation}
%
As anticipated above, this is truncatable, involving only the moments
$\mi=\int d\sig \rhsig \sig^i$ ($i=0\ldots 3$) of the density
distribution. Bartlett~\cite{Bartlett99} provided an elegant argument
why---at least within a virial expansion---such a moment structure of
the excess free energy for the hard sphere fluid should in fact be
exact.


For phase coexistence calculations we will also need to have a compact
expression for the excess free energy of the polydisperse hard sphere
{\em crystal}. This is not at all a trivial question. In principle,
the structure of a polydisperse crystal could be rather complex, with
different sites inside the crystalline unit cell occupied
preferentially by particles with different ranges of diameters. The
system would then effectively be an ordered solid solution (see
e.g.~\cite{CotParVegMon96,CotMon95}). Most theoretical work makes the
simplifying assumption that one has a substitutionally disordered
solid, where crystal sites are assumed to be occupied equally likely
by particles of any diameter (see e.g.~\cite{BarBauHan86,KraFre91}).

A simple-minded but popular approach to estimating the free energy is
cell theory, first introduced by Kirkwood~\cite{kirkwood50} and
widely used since (see e.g.~\cite{Sear98}): particles are treated as
independent but confined to an effective cell formed by their
neighbours%
. However, it is clear that for a polydisperse system this is unlikely
to be a useful approximation. For example, the cells of the model
would have to be made large enough to accommodate the particles with the
largest diameter, even if the fraction of such particles is very small.

We follow instead the more quantitative, ``geometric'' approach
proposed by Bartlett~\cite{Bartlett99,Bartlett97}.  He assumed that
the excess free energy of the solid depends on the same moments $\mze,
\ldots, \mth$ as that of the fluid. This can be motivated from scaled
particle theory~\cite{ReiFriLeb59,LebHelPar65}, which suggests that
the excess chemical potential, $\muexsig$, of spheres of
diameter $\sig$ is given by a cubic polynomial in $\sig$
\begin{equation}
\muexsig = \muze^{\rm ex} + \muon^{\rm ex} \sig + \mutw^{\rm ex} \sig^2 +
\muth^{\rm ex} \sig^3
\label{eq:muex_spt}
\end{equation}
The coefficients $\muze^{\rm ex}$ and $\muth^{\rm ex}$ can be
determined from the Widom insertion principle~\cite{Widom63}. The
latter can be stated as saying that $\exp(-\muexsig)$ is the ratio of
the (excess parts of the) partition functions for $N+1$ and $N$
particles, where the added particle has diameter $\sig$.
(Equivalently the excess chemical potential may be interpreted as the
work of inserting an $(N+1)$-th hard sphere of diameter $\sig$ into a system
of $N$ spheres.) In a system with purely hard interactions, this
implies that $\muexsig$ is positive and an increasing function of
$\sig$. For large $\sig$, the presence of the added particle
effectively just reduces the volume available to the $N$ others,
giving $\muexsig\approx P\sig^3$ ($=\tilde{P}(\pi/6)\tilde{\sig}^3$ in
dimensional units), hence $\muth^{\rm ex}=P$. For small $\sig$, one
notes that the ratio of the (excess) partition functions is also the average
Boltzmann factor of the added particle, the average being over the
Boltzmann distribution of the $N$-particle system. In the hard sphere
case, $\exp(-\muexsig)$ is thus the probability of being able to
insert a particle without overlap.  In the limit of vanishing particle
this probability is $1-\phi$, giving $\mu^{\rm ex}(\sig\to
0)=\muze^{\rm ex}=-\ln(1-\phi)$.

One now notes that (\ref{eq:muex_spt}) implies that the excess free
energy can only depend on the moments $\mze, \ldots,
\mth$. Indeed, from the definition of the excess chemical potentials
and with the dependence of the excess free energy on $\rhsig$
expressed through a (possibly infinite) set of moments $\mi$,
\[
\muexsig = \frac{\delta \fex}{\delta\rhsig} = \sum_i \muex_i \sig^i,
\qquad
\muex_i = \frac{\partial \fex}{\partial \mi}
\]
A comparison with the form (\ref{eq:muex_spt}) of the excess chemical
potentials reveals that $\fex$ can only depend on $\mze,\ldots,
\mth$, as claimed. The same is then true also for the $\muex_i = \partial
\fex/\partial \mi$ ($i=0,\ldots,3$), which are recognised as excess moment
chemical potentials.
The excess free energy of a {\em polydisperse} hard sphere mixture can
thus be deduced from that of any other mixture which is equivalent in
the sense of having the {\em same} $\mze, \ldots, \mth$. These
moments determine the number density along with the basic geometric
properties of mean particle diameter, surface area and volume.
The simplest mixture with a finite number of species that can match
any given $\mze, \ldots, \mth$ is a bidisperse one. Indeed, this has
four degrees of freedom, namely, the number densities and particle
diameters of the two species.
We can therefore identify the excess free energy of a polydisperse
hard sphere solid with that of the equivalent bidisperse system. For
the latter, we follow Bartlett in using the fits to the simulation
data of Kranendonk {\em et al}~\cite{KraFre91}. Because these data are
obtained for an fcc substitutionally disordered crystal, an implicit
assumption is that the polydisperse crystal will have the same
structure.

There is a difficulty in Bartlett's approach with the determination of
the excess moment chemical potentials $\muze^{\rm
ex},\ldots,\muth^{\rm ex}$. He fixed $\muze^{\rm ex}$ and $\muth^{\rm
ex}$ to the exact results derived from the Widom insertion principle,
$\muze^{\rm ex}=-\ln(1-\mth)$ and $\muth^{\rm ex}=P$. The remaining
two excess moment chemical potentials, $\muon^{\rm ex}$ and
$\mutw^{\rm ex}$, can then be found from the bidisperse simulation
data, by requiring $\muexsig$ at the diameters of the small and large
spheres to agree with the simulated excess chemical potentials of the
two species. However, because of the approximate character of the
excess free energy, the $\muexsig$ derived by this route do not obey
the thermodynamic consistency requirement
$\delta\muexsig/\delta\rh(\sig') = \delta\muex(\sig')/\delta\rhsig$,
which corresponds to $\partial \mu_i^{\rm ex}/\partial\rh_j = \partial
\mu_j^{\rm ex}/\partial\rh_i$ for the excess moment chemical
potentials. To avoid this in our study, we assign the latter by
explicitly evaluating the derivatives of the excess free energy,
$\muex_i = \partial \fex/\partial\mi$%
. Thermodynamic
consistency is then automatic. The price we pay is that our $\muexsig$
no longer has the theoretically expected asymptotic behaviour for
$\sig\to0$ and $\sig\to\infty$. This means that we have to restrict
use of our solid free energy to relatively narrow diameter
distributions, as discussed in more detail below.

\section{\label{sec:3hsNumericss}Numerical method}
\label{sec:MFE}

\subsection{Moment free energy}

 
Our computational approach for determining the phase behaviour of
polydisperse hard spheres is based on the moment free energy (MFE)
method.
We give a brief outline here; details can be found
in~\cite{SolWarCat01,Warren98,SolCat98,Sollich02}. Recall first the
phase equilibrium conditions for coexistence of $p$ phases, in a
system described by a truncatable free energy. By definition, the
excess free energy 
then depends on a finite set of $M$ (generalised) moments $\mi =
\int\!  d\sig\,\rhsig \wi$ defined by weight functions $\wi$; above we
had $\wi=\sig^i$. In coexisting phases, the chemical potentials
$\mu(\sig)$ and pressure $P$ must be equal. The former are, by
differentiation of (\ref{eq:fre_ene}),
\be
\musig = \frac{\delta f}{\delta\rhsig} = \ln\rhsig + 
\sum_i \muex_i w_i(\sig)
\label{musig}
\ee
with $\muex_i = {\partial\fex}/{\partial\mi}$ as before.
The pressure is given by the Gibbs-Duhem relation
\be
P = 
-f + \intsig\musig\rhsig = \rh_0 - \fex + \sum_i \muex_i \mi
\label{P_trunc}
\ee
To the conditions of equality of chemical potentials and pressure we
need to add the requirement of conservation of particle number for
each species $\sig$, which reads
\be
\sum_\alpha v\pa \rh\pa(\sig) = \rhzsig
\label{particle_cons}
\ee
where $\alpha=1,\ldots,p$ labels the phases and $v\pa$ is the fraction
of the system volume occupied by phase $\alpha$. One then finds
from equality of the $\musig$, Eq.~(\ref{musig}), together with
particle conservation (\ref{particle_cons}), that the density
distributions in coexisting phases can be written as
\be
\rh\pa(\sig) = \rhzsig \, \frac{\exp\left[\sum_i \lambda_i\pa\wi\right]}
{\sum_\gamma v^{(\gamma)} \exp\left[\sum_i \lambda_i^{(\gamma)}\wi\right]}
\label{rhoalsig_trunc}
\ee
Here the $\lambda_i\pa$ must obey
\be
\lambda_i\pa = -\mu_i^{(\alpha),{\rm ex}} + c_i
\label{lambda_i_exact}
\ee
and the $c_i$ are undetermined constants that do not affect the
density distributions (\ref{rhoalsig_trunc}). One can fix them e.g.\
by requiring all the $\lambda_i\pa$ in one of the phases to be zero.
A little reflection then shows that\eq{lambda_i_exact} together with
$\sum_\alpha v\pa=1$ and the equality of the pressures\eq{P_trunc} in
all phases give a closed system of nonlinear equations for the
$p(M+1)$ variables $\lambda_i\pa$ and $v\pa$. A solution can thus, in
principle, be found by a standard algorithm such as
Newton-Raphson. Generating an initial point from which such an
algorithm will converge, however, is still a nontrivial problem,
especially when more than two phases coexist and/or many moments $\mi$
are involved. Furthermore, the nonlinear phase equilibrium equations
permit no simple geometrical interpretation or qualitative insight
akin to the construction of phase diagrams from the free energy
surface of a finite mixture.

The moment free energy addresses these two disadvantages. To construct
it, one starts by modifying the free energy decomposition\eq{eq:fre_ene}
to
\be
f = \intsig \rhsig \left[\ln \frac{\rhsig}{\prior}
-1\right] + \fex(\{\mi\})
\label{free_en_decomp}
\ee
In the first (ideal) term, a normalising factor $R(\sigma)$ has been
included inside the logarithm. This has no effect on the exact
thermodynamics because it contributes only terms linear in $\rhsig$,
but will play a central role below. One can now argue that the most
important moments to treat correctly in the calculation of phase
equilibria are those that actually appear in the excess free energy
$\fex(\{\mi\})$. Accordingly, one imposes particle
conservation\eq{particle_cons} only for the $\mi$, but allows it to be
violated in other details of the density distribution $\rhsig$ which
do not affect the $\mi$. These ``transverse'' degrees of freedom are
instead chosen to minimise the free energy\eq{free_en_decomp}, and
more precisely its ideal part since the excess contribution is a
constant for fixed values of the $\mi$. This minimisation gives
\be
\rhsig=\prior\exp\left[\sum_i \lambda_i\wi\right]
\label{family}
\ee
where the Lagrange multipliers $\lambda_i$ are chosen to give the
desired values of the moments
\be
\mi = \intsig\wi\,\prior\exp\left[\sum_j \lambda_j w_j(\sig)\right]
\label{moments_from_lambda}
\ee
The corresponding minimum value of $f$ as given in\eq{free_en_decomp}
then defines the {\em moment free energy} (MFE)
\be
\fmom(\{\mi\}) = \left(\sum_i \lambda_i\mi - \mze \right) + \fex(\{\mi\})
\label{fmom}
\ee
Since the Lagrange multipliers are (at least implicitly) functions of
the moments $\mi$, the MFE depends only on the $\mi$. These can now be
viewed as densities of ``quasi-species'' of particles, allowing for
example the calculation of moment chemical
potentials~\cite{SolWarCat01}
\be
\mu_i= \frac{\partial\fmom}{\partial\mi} =
\lambda_i + \frac{\partial\fex}{\partial\mi} = \lambda_i + \muex_i
\label{mom_chem_pot}
\ee
and the corresponding pressure $P = \sum_i \mu_i \mi - \fmom$ which
turns out to be identical to the exact expression\eq{P_trunc}. A
finite-dimen\-sio\-nal phase diagram can thus be constructed from
$\fmom$ according to the usual tangency plane rules, ignoring the
underlying polydisperse nature of the system.  Obviously, though, the
results now depend on $R(\sig)$.  To understand its influence, one
notes that the MFE is simply the free energy of phases in which the
density distributions $\rhsig$ are of the form\eq{family}. To ensure
that the parent phase is contained in the family, one normally chooses
its density distribution as the prior, $R(\sig)=\rhzsig$; the MFE
procedure will then be exactly valid whenever the density
distributions actually arising in the various coexisting phases are
members of the corresponding family
\be
\rhsig=\rhzsig\exp\left[\sum_i \lambda_i\wi\right]
\label{pfamily_precap}
\ee
It is easy to show from\eq{rhoalsig_trunc} that this condition holds
whenever all but one of a set of coexisting phases are of
infinitesimal volume compared to the majority phase.  Accordingly, the
MFE yields {\em exactly} the onset of phase of coexistence,
conventionally represented via cloud and shadow curves (see
below). Similarly, one can show that spinodals and critical points are
found exactly~\cite{SolWarCat01}.

For coexistences involving finite amounts of different phases the
MFE only gives approximate results, since different
density distributions from the family\eq{pfamily_precap},
corresponding to two (or more) phases arising from the same parent
$\rhzsig$, do not in general add to recover the parent distribution
itself. Moreover, from Gibbs' phase rule, a MFE
depending on $M$ moments will not predict more than $M+1$ coexisting
phases, while we know that a polydisperse system can in principle
separate into an arbitrary number of phases. Both of these
shortcomings can be overcome by including extra moments within the
MFE. By choosing the weight
functions of the extra moments adaptively, the properties of the
coexisting phases can then be predicted with in principle arbitrary
accuracy~\cite{ClaCueSeaSolSpe00,SolWarCat01}. Importantly for us, the
results can in fact be used as initial points from which a solution of
the exact phase equilibrium problem can be converged
successfully~\cite{SpeSol02,SpeSol03a}. This is the technique that we
use here. Once a phase split for a given parent distribution $\rhzsig$
has been found, care needs to be taken to check that it is globally
stable, i.e.\ that no phase split of lower free energy
exists~\cite{SolWarCat01}. Adopting this procedure, we are able to
calculate coexistence of up to five phases, which so far has been
possible only for much simpler free energies depending on a single
density moment (see e.g.~\cite{SolWarCat01}).

\subsection{Implementation}

We focus below on parent distributions with unit mean particle
diameter $\overline\sig$; any other choice could be absorbed into the
unit length $\unil$. For small polydispersity $\pol$, the standard
moments $\mi = \intsig\rhsig \sig^i$ then become very close to each
other, and in fact strictly identical in the limit $\pol\to 0$. This
causes numerical difficulties, and we therefore work instead with the
centred moments $\rh\ctr_i=\intsig\rhsig\cwi$ which remain distinct
even for small $\pol$. The factor $\pol_0$ is included to ensure that
the moments are all of comparable magnitude. We therefore choose it in
the middle of the range of polydispersities $\pol$ that we study, with
typically $\pol_0=0.05$. 
%
%
The centred moments are obviously linearly related to the conventional
ones, e.g.\ $\rh\ctr_1=(\mon-\mze)/\pol_0$. The BMCSL and solid free
energies can therefore readily be re-expressed in term of the centred
moments. Because the transformation between the two sets of moments is
linear, the corresponding sets of excess moment chemical potentials
$\muex_i = \partial\fex/\partial\mi$ are also linearly related and
easily converted into each other.
%


We combine the fluid and solid branches of our excess free energy by
simply taking the minimum for a given set of moments. Some care is
needed here: because the solid free energy is derived from fits to
simulation data for bidisperse systems (see above
), we expect it to be reliable only in the region spanned by the
simulations~\cite{KraFre91}. The smallest diameter ratio
investigated in the simulations
is $\alpha=0.85$. The maximum polydispersity that can be reached in a
bidisperse system for this diameter ratio is $\pol =
(\alpha+1/\alpha-2)^{1/2}/2 \approx 0.08$. We therefore restrict use
of the solid free energy to phases with polydispersity below this
value. Reassuringly, we will see below that all solid phases occurring
in equilibrium phase splits are well below this threshold.

A further constraint on the use of the solid free energy arises from
the fact that, as explained above, our excess chemical potentials
$\muexsig$ do not have the correct limiting behaviour predicted from
the Widom insertion principle for $\sig\to 0$ and $\sig\to
\infty$. Physically, this is again plausible because we are
extrapolating to sphere diameters far from the mean of the
distribution, and therefore far from the regime where the simulation
data will be reliable. We will therefore always work with diameter
distributions with hard cutoffs either side of the mean so that the
behaviour of $\muexsig$ for very small or large $\sig$ never comes
into play. Finally, we also restrict the volume fractions for the
solid branch of the free energy to $0.494 \leq \phi \leq 0.74$, which
are respectively the smallest and largest $\phi$ for which
monodisperse hard spheres at equilibrium exhibit a crystalline solid
phase.

%

\section{Phase behaviour}
\label{sec:phasediag}

We now describe our results for the overall phase behaviour of
polydisperse hard spheres. Our numerical work requires a choice to be
made for the parental diameter distribution. We focus mostly on a
triangular distribution, where 
$n^{(0)}(\sig) = \rhzsig/\mze^{(0)}$ is given by
\[
n^{(0)}(\sig) = \frac{1}{w^2} \left\{
\begin{array}{llrcl}
\sig-(1-w) \quad & \mbox{for} & 1-w & \leq \sig \leq & 1 \\
(1+w)-\sig \quad & \mbox{for} & 1 & \leq \sig \leq & 1+w
\end{array}\right.
\]
whose width parameter $w$ is related to the polydispersity by
$w=\sqrt{6}\,\pol$. For the moderate values of $\pol$ of interest here
one expects other distribution shapes to give qualitatively similar
results, based on the intuition that for narrow size distributions
$\pol$ is the key parameter controlling the phase
behaviour~\cite{Pusey87}. To verify this, we also consider briefly a
Schultz parent distribution, $n^{(0)}(\sig) \propto
%
%
\sig^z e^{-(z+1)\sig}$, cut off outside the range
$\sig\in[0.8,1.2]$. For a narrow distribution, i.e.\ large $z$, where
the cutoffs are
unimportant, the polydispersity is then related to the parameter $z$
by $\pol^2=1/(z+1)$ and the mean diameter is unity as before.

\subsection{Onset of phase coexistence}

\begin{figure}
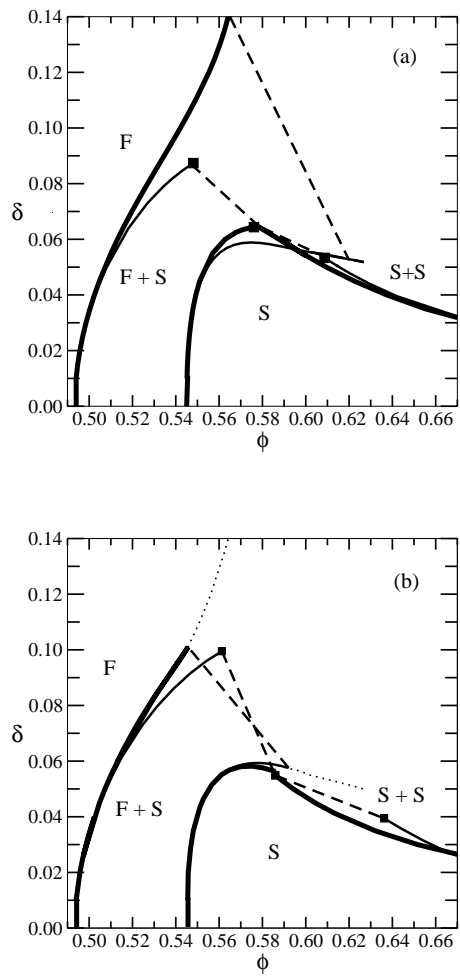

\begin{minipage}{.45\textwidth}
    \begin{center}  
	\includegraphics[width=6cm]{./cloud_shadow_triangular.eps}%
    \end{center}
\end{minipage}

\vspace{28pt}

\begin{minipage}{.45\textwidth}
  \begin{center}  
      \includegraphics[width=6cm]{./cloud_shadow_schultz.eps}%
  \end{center}  
\end{minipage}
\caption{Cloud curves (thick) and shadow curves (thin),
	for polydisperse hard spheres with a triangular (a) and
	Schultz (b) diameter distribution. The curves show
	polydispersity $\pol$ versus volume fraction $\phi$ for the
	cloud and shadow phases; dashed lines link sample cloud-shadow
	pairs.  The solid (S) cloud curve has two branches, with onset
	of F-S and S-S coexistence at low and high volume fractions,
	respectively. Where they meet, a triple point occurs; squares
	mark the cloud phase and the two coexisting shadows there. In
	the Schultz plot, the dotted lines indicate the expected
	continuations of the fluid cloud and corresponding shadow
	curve beyond the region where our numerical methods work
	reliably. \label{fig:cloud_shadow} }
\end{figure}

The most basic question we can ask about phase behaviour regards the
onset of phase separation coming from single-phase regions. Increasing
the volume fraction $\phi$ of the parent at given polydispersity
$\pol$, a single-phase fluid (F) will first separate into coexisting
fluid and solid (S) phases at the so-called cloud point. The locus of all
cloud points in the ($\phi$,$\delta$)-plane defines the fluid cloud
curve. The incipient phase corresponding to the cloud point is called
the ``shadow'' solid; its properties define the solid shadow curve.
These curves are shown in Fig.~\ref{fig:cloud_shadow}~(a) for a
triangular parent distribution. An important feature is that the fluid
cloud curve continues throughout the whole range of polydispersities
that we can investigate numerically: even at $\pol=0.14$, a hard
sphere fluid will eventually split off a solid on compression.
This is in marked contrast to the phase diagram of~\cite{BarWar99} as
sketched in Fig.~\ref{fig:pha_dia_ske}. The key difference is that our
analysis accounts fully for fractionation: Fig.~\ref{fig:cloud_shadow}
shows that the coexisting shadow solid always has a relatively modest
polydispersity, with $\pol$ never rising above $0.06$ even when the
cloud fluid has $\pol=0.14$. This fractionation effect prevents the
convergence of the solid and fluid phase boundaries which a theory
with fixed $\pol$~\cite{BarWar99} predicts, along with the resulting
re-entrant melting (Fig.~\ref{fig:pha_dia_ske}). These findings are in
qualitative accord with Monte Carlo simulations for the simpler case
of fixed chemical potentials~\cite{BolKof96,BolKof96b,KofBol99},
discussed in detail in
Section~\ref{sec:3hsComparisonwithBolhuisandKofkeresults} below. The
results imply, in particular, that the terminal polydispersity $\polt$
cannot be defined as the point beyond which a fluid at equilibrium
will no longer phase separate; $\polt$ only makes sense as the maximum
polydispersity at which a single solid phase can exist (see below).

\begin{figure}
  \begin{center}
      \includegraphics[width=8cm]{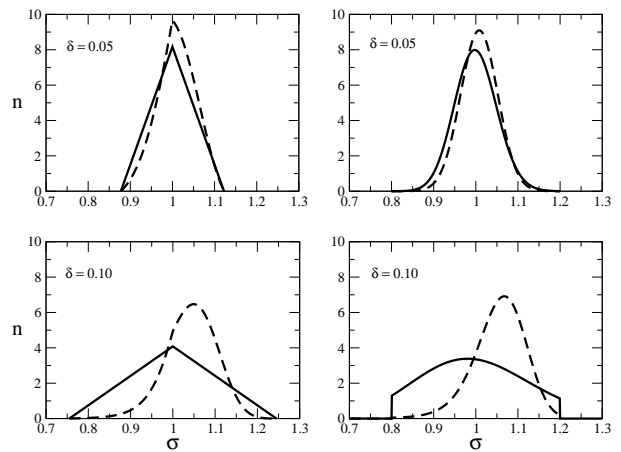}
      \caption{Normalised size distributions $\nsig = \rhsig/\mze$ for
      the coexisting fluid and shadow phases at the fluid cloud point,
      for parent distributions with polydispersities $\pol=0.05$ (top)
      and $\pol=0.10$ (bottom) and triangular (left) and Schultz
      (right) shape.
      Solid lines show the cloud fluid, which is identical to the
      parent, and dashed lines the shadow solid. 
\label{fig:fs_clo_poi_dis}
}
  \end{center}
\end{figure}

The fractionation effects described above can be seen more explicitly
by comparing the (normalised) diameter distributions of the fluid
cloud and solid shadow phases, as displayed in
Fig.~\ref{fig:fs_clo_poi_dis} for (parental) polydispersity
$\pol=0.05$ and $\pol=0.10$. At the cloud point the size distribution
in the fluid coincides with the parent distribution as it must. The
distribution in the coexisting shadow solid, on the other hand,
deviates increasingly from that of the parent as the parental $\pol$
increases. In particular, the solid contains predominantly the larger
particles and has a rather more narrow spread of sizes, consistent with
the small solid polydispersities found above. We will see shortly that these
properties are rather generic and persist inside the coexistence
region.

We now assess the effect of the shape of the particle size distribution on
these results. Fig.~\ref{fig:cloud_shadow} shows that the fluid cloud
and solid shadow curves are qualitatively and even quantitatively very
similar for the triangular and Schultz distributions. (Numerically, we
can only reach $\pol=0.10$ for the latter, but have no reason to
expect that this is a physical feature and indicate the expected
continuation of the curves by dotted lines.)
Fig.~\ref{fig:fs_clo_poi_dis} (right) demonstrates that the
qualitative features of the fractionation behaviour are also the same
between the two distributions, consistent with our intuition that
variations in the shape of the parental size distribution have, for
given $\pol$, only a minor effect.

We next consider the onset of phase separation coming from the
single-phase solid, which defines the solid cloud curve and
corresponding shadow curve. Initially we focus again on the triangular
size distribution. Fig.~\ref{fig:cloud_shadow} (a) shows that a
decrease in density at low polydispersities leads to conventional
fluid-solid phase separation. At higher $\pol$, however, the solid
cloud curve acquires a second branch at higher densities. This is
broadly analogous to the re-entrant phase boundary found
in~\cite{BarWar99}, but with the crucial difference that the system
phase separates into two solids rather than a solid and a fluid. The
two branches meet at a triple point. Here the solid cloud phase
coexists with {\em two} shadow phases, one fluid and one solid, as
marked by the squares in Fig.~\ref{fig:cloud_shadow} (a). The triple
point is located at $\pol\approx 0.07$; since it is at the maximum of
both branches of the solid cloud curve, this value also gives the
terminal polydispersity beyond which solids with triangular diameter
distribution are unstable against phase separation.


\begin{figure}
 \begin{center}
      \includegraphics[width=8cm]{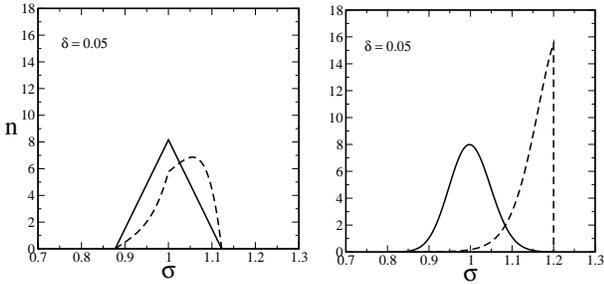}%
      \end{center}
\caption{Normalised size distributions $\nsig$ of solid
      cloud and shadow phases, on the high-density branch of the solid
      cloud curve at $\pol=0.05$. Left: triangular parent; right:
      Schultz parent. The solid lines show the cloud phase, the dashed
      lines the shadow. Note the strong size fractionation effects. 
\label{fig:ss_clo_poi_dis} }
\end{figure}
Fig.~\ref{fig:ss_clo_poi_dis} (left) displays the diameter
distributions for the cloud and shadow solids, at $\pol=0.05$ on the
high-density branch of the solid cloud curve. In comparison with
Fig.~\ref{fig:fs_clo_poi_dis}, what is striking is that the
fractionation effects at the onset of solid-solid coexistence are much
stronger than for fluid-solid phase separation at the same $\pol$.
This is consistent with physical intuition. The fluid-solid phase
separation exists even in the monodisperse limit. The presence
of polydispersity acts as a small perturbation to this transition, certainly
at low $\pol$, so that fractionation effects can be viewed as
incidental. Solid-solid phase separation, on the other hand, is driven
by polydispersity and could not take place without fractionation.

We compare again at this stage with the results for the Schultz parent
distribution. Fig.~\ref{fig:cloud_shadow} (b) shows that the cloud
and shadow curves look qualitatively similar to the triangular
case. Quantitatively, the low-density branch of the solid cloud curve
now has a maximum, giving the terminal polydispersity as $\polt\approx
0.06$. The triple point is at slightly smaller $\pol$, and the whole
high-density branch of the solid cloud curve -- which describes the
onset of solid-solid phase separation -- is shifted to smaller $\pol$
compared to the triangular parent case.  Fig.~\ref{fig:ss_clo_poi_dis}
(right) shows the diameter distributions for the cloud and shadow
solids, at the onset of phase separation at $\pol=0.05$. Compared to
the triangular parent, the fractionation effects are now even
stronger. In fact, the size distribution of the shadow solid continues
to increase towards larger sphere diameters $\sig$ and is terminated
only by the hard cutoff at $\sig=1.2$ which we impose in the Schultz
case; note that in the {\em cloud} solid (solid line) this cutoff is
hardly discernible. In the triangular case, there is no sharp cutoff
effect on the shadow solid: the form of the parent forces all size
distributions to drop to zero continuously at the upper end.


\begin{figure}
 \begin{center}
      \includegraphics[width=8cm]{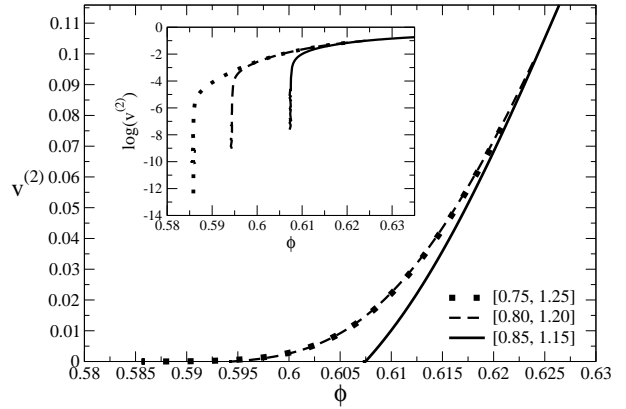}%
      \end{center}
\caption{
Fractional system volume occupied by the newly forming solid,
$v^{(2)}$, versus the parent volume fraction, $\phi$, for cutoffs
imposing three different ranges of particle sizes $\sigma$ as indicated in the
legend. Once enough of the new solid phase exists (above $v^{(2)}\approx
0.08$) the behaviour is essentially cutoff-independent. The cloud
point, on the other hand, where $v^{(2)}$ extrapolates to zero,
depends strongly on the cutoff; this is more clearly visible in the
linear-log plot in the inset.
\label{fig:Schultz_cutoff} }
\end{figure}

The above observations for the Schultz parent suggest an analogy with
recent results for isotropic-nematic phase separation in hard rod-like
particles~\cite{SpeSol03a,SpeSol03b}. For sufficiently wide rod length
distributions, one observes there that the shadow nematic phase can
become dominated by the longest rods in the system, i.e.\ those with
lengths near the cutoff, even though these make up only a small
fraction of the parent distribution. Such cutoff effects are important
only near the cloud point: as soon as the new phase occupies a nonzero
fraction of the overall system volume, particle conservation prevents
it from containing an atypically large number of long rods.  To test
whether we have a similar situation for the onset of solid-solid
separation from a Schultz parent, we have varied the cutoff on the
sphere diameters. Fig.~\ref{fig:Schultz_cutoff} plots the fractional
system volume $v^{(2)}$ occupied by the new solid against the parent
colloid volume fraction. We observe that $v^{(2)}$ is indeed
cutoff-independent well inside the coexistence region, where it is
non-negligible. The position of the cloud point itself, on the other
hand, where $v^{(2)}$ extrapolates to zero, is strongly
cutoff-dependent. We conclude, therefore, that at the onset of
solid-solid coexistence from a Schultz parent with $\pol=0.05$ the
shadow solid is cutoff-dominated, in analogy with the shadow nematics
in the Onsager model of long hard rods~\cite{SpeSol03a,SpeSol03b}.

One may wonder whether the cutoff effects described above are an
artefact of the approximate nature of our excess chemical potentials
for the polydisperse solid. We cannot give a definitive answer to this
question here, but suggest that such effects might in fact be
expected. From (\ref{musig}), equilibrium of chemical potentials
between the cloud (parent) solid $\rhzsig$ and the shadow solid
$\rh^{(2)}(\sig)$ implies
\begin{equation}
\rh^{(2)}(\sig) = \rhzsig \exp(-\Delta\muexsig)
\label{Schultz_intuition}
\end{equation}
with $\muexsig = \sum_{i=0}^3 \Delta\muex_i \sig^i$ and
$\Delta\muex_i$ the differences in the excess moment chemical
potentials between the shadow and cloud phases. The Widom insertion
principle tells us that $\muth^{\rm ex}$, being equal to the pressure,
is equal in the two phases. Thus $\Delta\muexsig$ should generically
be a quadratic polynomial in $\sig$. If the $\sig^2$-term has a
negative coefficient, then from (\ref{Schultz_intuition}) it will
overwhelm the exponential tail of the Schultz parent $\rhzsig$.
The shadow density distribution $\rh^{(2)}(\sig)$ then increases
strongly at large $\sig$ so that its properties will be dominated by
the presence of 
any cutoff. In fact, this argument suggests that the same could happen
even with a (sufficiently polydisperse) Gaussian parent. Only a stronger decay,
say $\rhzsig \sim \exp(-\sig^a)$ with $a>2$, could definitely prevent
cutoff effects on the shadow solid. This question deserves further
study, but would require more accurate excess chemical
potentials for polydisperse solids -- and over a larger range of
sphere diameters -- than we currently have at our disposal.

So far we have investigated the global stability of single phases,
i.e.\ the stability against macroscopic phase separation. One can also
ask about {\em local} stability of the phases, i.e.\ the spinodal
points. Since our free energy is spliced together from separate
fluid and solid branches, we cannot investigate instability to
fluid-solid separation. The stability of single-phase fluids against
fluid-fluid demixing has been studied by Warren~\cite{Warren99} and
Cuesta~\cite{Cuesta99}. They found, using the BMCSL free energy, that
spinodal instabilities do indeed occur, but only for very broad
diameter distributions such as log-normals with $\pol$ above $\approx
2.5$, or bimodal distributions with a wide size disparity between the
larger and smaller spheres. At the modest values of $\pol$ that
concern us here, such instabilities do not occur.
%
\begin{figure}
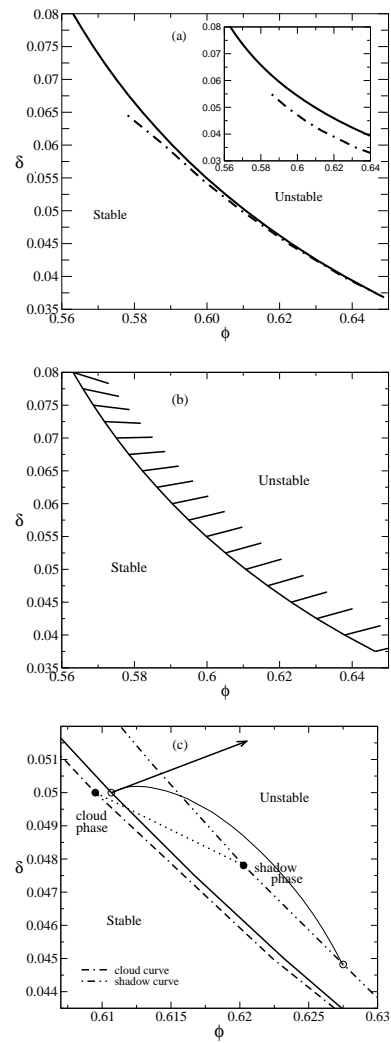

\begin{minipage}{.32\textwidth}
    \begin{center} 
	 \includegraphics[width=5cm]{./cloud_shadow_triangular_spinodal.eps}	 
    \end{center}
\end{minipage}

\vspace{8pt}

\begin{minipage}{.32\textwidth}
    \begin{center} 
	 \includegraphics[width=5cm]{./spinodal.eps} 
    \end{center}
\end{minipage}

\vspace{8pt}

\begin{minipage}{.32\textwidth}
  \begin{center} 
    \includegraphics[width=5cm]{./spinodal_minimizer.eps}
  \end{center}  
\end{minipage}
\caption{Spinodal instability of the polydisperse hard sphere crystal against
solid-solid demixing, in the volume fraction -- polydispersity plane
($\phi$,$\pol$). (a) Spinodal (solid) and cloud curve (dash-dotted) for
triangular (main graph) and Schultz (inset) size distributions.  (b)
The line segments on the spinodal line indicate (for the triangular
case) the direction of the unstable fluctuations. (c) Comparison of
instability direction (arrow), path to the ``locally optimal'' phase
(solid line and empty circles), and cloud and shadow solids at the
same parent polydispersity (full circles connected by dotted line).
\label{fig:spinodal}
}  
\end{figure}
%
It thus remains to study spinodal instabilities of the polydisperse
crystal against solid-solid demixing. The fact that the solid cloud
curve has a branch showing solid-solid phase separation already
suggests that such instabilities should be present. Indeed, Bartlett
found a solid-solid spinodal~\cite{Bartlett00}, though with a
thermodynamically inconsistent assignment of the excess chemical
potentials (see Section~\ref{sec:3hsfreeenergychoice}). Within the MFE
the criterion for the spinodal takes its usual form~\cite{SolWarCat01}:
it is the point where the determinant of the curvature matrix of the
moment free energy, $\partial^2 \fmom/(\partial \mi\partial \rh_j) = \partial
\mu_i/\partial \rh_j$, first vanishes. The zero eigenvector of the
matrix at this point gives the instability direction.
Using this criterion, we find the results in
Fig.~\ref{fig:spinodal}~(a). 
The single-phase solid is always stable at modest densities or
polydispersities -- the spinodal determinant is positive here -- but
can become unstable at larger $\phi$ and $\pol$. With growing $\pol$,
this instability affects a wider and wider range of $\phi$. The figure
also shows that the spinodal for a triangular size distribution is
very close to the cloud curve for the onset of solid-solid phase
separation: past the cloud point, a single-phase solid very quickly
becomes locally unstable. For a Schultz distribution, on the other
hand, cloud curve and spinodal are well separated as can be seen in
the inset. This reinforces our above discussion of cutoff effects: the
latter favour an earlier onset of phase separation,
cf.~Fig.~\ref{fig:Schultz_cutoff}. The spinodal condition, on the
other hand, is known on general grounds (see e.g.~\cite{SolWarCat01})
to depend only on the moment densities of the parent, up to $\sig^6$
for our excess free energy involving moments up to $\sig^3$. Since
these parent moments are almost cutoff-independent, so is the spinodal
curve. (This insensitivity of the location of the spinodal is
confirmed by the fact that the spinodal curves for the triangular and
Schultz cases would be essentially indistinguishable on the scale of
Fig.~\ref{fig:spinodal} (a).)

We now turn to the nature of the spinodal instability, focusing on
the case of a triangular size distribution. This can be quantified by
projecting the instability direction at the spinodal into the
($\phi$,$\pol$)-plane.
The results are indicated by the line segments on the spinodal line in
Fig.~\ref{fig:spinodal} (b). Bartlett found instability directions
which affected only the polydispersity $\pol$ while leaving $\phi$
essentially unchanged~\cite{Bartlett00}, which would correspond to
vertical lines in the plot. By contrast, our analysis shows that the
instability actually affects both $\phi$ and $\pol$, with relative
changes that are of the same order of magnitude. This is consistent
with the properties of coexisting solids discussed in
Sec.~\ref{sec:3hsPhaseDiagram} below, which exhibit a strong
correlation between $\phi$ and $\pol$.

More puzzling is that, in Fig.~\ref{fig:spinodal} (b), the
instability directions at low $\pol$ indicate a tendency towards the
formation of a {\em more} polydisperse solid. This appears
counter-intuitive at first: solid-solid phase
separation is driven by fractionation and so one expects a preference
for smaller rather than 
larger $\pol$. Also, the proximity of the spinodal to the cloud curve
suggests that the spinodal instability direction should be similar to
the direction connecting the cloud solid and the coexisting shadow.
From Fig.~\ref{fig:cloud_shadow}, the instability should therefore
point towards larger $\phi$ and, again, smaller $\pol$.

To understand this apparent paradox we consider in more detail the
``shape'' of the MFE $\fmom$ at the spinodal, as a function of the
moments $\rho_0$, \ldots, $\rho_3$. It is useful to subtract the
tangent plane to $\fmom$ at the parent phase; the resulting tangent
plane distance (tpd) differs from $\fmom$ only via constant and linear
terms in the $\rho_i$. A stable parent is then a local minimum of the
tpd, at ``height'' ${\rm tpd}=0$, and any phases coexisting with the
parent (e.g.\ the shadow phase(s) for a parent at its cloud point)
would have the same property. Now, as the spinodal is approached, the
curvature of the tpd around the parent vanishes in one direction and a
``path'' towards lower, negative, values of the tpd appears; the
spinodal instability indicates the initial direction of this path. To
establish where this path leads it makes sense to follow it to the
nearest ``locally optimal'' phase, i.e.\ the nearest local minimum of
the tpd. If this path is curved in the $(\rho_0,\ldots,
\rho_3)$-space, its initial direction will not necessarily capture
the properties of the end point, i.e.\ the locally
optimal phase. This is the origin of the counter-intuitive instability
directions that we observe. A specific example is shown in
Fig.~\ref{fig:spinodal} (c): the path to the locally optimal phase
first moves to higher $\pol$, consistent with the spinodal instability
direction, but the locally optimally phase ends up having a {\em
smaller} polydispersity $\pol$ than the parent phase. It also has
a larger volume fraction $\phi$, and the change from the unstable
parent to the locally optimal phase is in a direction comparable to that
between cloud and shadow, in line with the intuition discussed above.

\subsection{Phase diagram}
\label{sec:3hsPhaseDiagram}

Having clarified the {\em onset} of phase separation in polydisperse
hard spheres, we next consider the behaviour inside the coexistence
region. We have already established that, apart from possible cutoff
effects, the Schultz and triangular parent size distributions give
qualitatively similar results, and therefore restrict our attention to
the latter in the following. Overall features such as the topology of
the phase diagram should, at the low polydispersities $\pol$ of
interest here, be similar for other size distributions.

\begin{figure}
\begin{center}
\includegraphics[width=8.5cm]{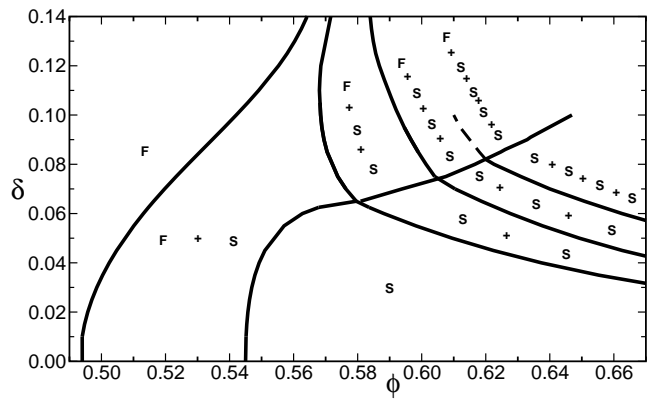}%
\caption{Full phase diagram for polydisperse hard spheres with a
triangular size distribution. In each region the nature of the
phase(s) coexisting at equilibrium is indicated (F: fluid, S:
solid). Dashed line: best guess for the phase boundary in the region
where our numerical data become unreliable.
From~\protect\cite{FasSol03}.
\label{fig:parent}
}
\end{center}
\end{figure}
Fig.~\ref{fig:parent} shows the full phase diagram for the triangular
parent distribution. In each region the nature of the phase(s)
coexisting at equilibrium is indicated. The cloud curves of
Fig.~\ref{fig:cloud_shadow} (a) reappear as the boundaries between
single-phase regions and areas of phase coexistence.
Starting from the onset of solid-solid separation and increasing density or
$\pol$, fractionation into multiple solids occurs. The overall shape
of the phase boundaries in this region is in good qualitative
agreement with the approximate calculations
of~\cite{Bartlett98}, see Fig.~\ref{fig:solids},
though as discussed below the details of the fractionation behaviour
are rather different. We find up to 4 coexisting
solids. At larger $\pol$ than we can
tackle numerically, phase splits into 5 or more solids would be expected
since each individual solid can only tolerate a finite amount of
polydispersity. However, from Fig.~\ref{fig:parent} such phase splits
would occur at increasing densities and eventually be limited by the
physical maximum volume fraction $\phi_c\approx 74\%$. Also,
at higher $\pol$ more complicated single-phase crystal structures,
with different lattice sites occupied preferentially by
(say) smaller and larger spheres, could appear and compete with the
substitutionally disordered solids we consider.

A feature of the phase diagram in Fig.~\ref{fig:parent} not predicted
in previous work is the coexistence of a fluid with multiple
solids. However, that a three-phase F-S-S region must occur was already
indicated by the triple point which we found earlier on the solid
cloud curves. As in the case of solid-solid phase splits,
coexistences involving more than two solids -- and a fluid --
then appear with increasing $\pol$.

\begin{figure}
\begin{center}
\includegraphics[width=8.5cm]{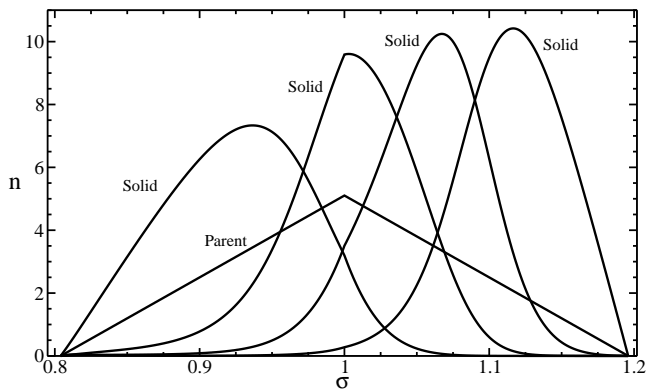}%
\caption{Normalised size distribution of four coexisting solid phases
obtained from a parent distribution with
$(\phi,\pol)=(0.63,0.08)$. From left to right, the solids have volume
fractions and polydispersities $(0.601,0.054)$, $(0.629, 0.046)$,
$(0.646, 0.040)$, $(0.663, 0.036)$. From~\protect\cite{FasSol03}.
\label{fig:solid_distribution}
}
\end{center}
\end{figure}
We consider the fractionation behaviour in the multiphase regions more
systematically in the next section. Before doing so, a few qualitative
statements are in order. In Fig.~\ref{fig:solid_distribution} we show
a sample plot of the normalised diameter distributions
$\nsig=\rhsig/\mze$ of four coexisting solids. This shows that
fractionated solids do not, as one might naively
assume~\cite{Bartlett98}, split the diameter range of the parent
evenly among themselves. The polydispersities of the coexisting phases
are in fact rather different; in Fig.~\ref{fig:solid_distribution}
they range from $\pol=0.036$ to $0.054$ for a parent with $\pol=0.08$.
There is in fact a strong correlation between the polydispersity of a
fractionated solid and its volume fraction: solids with lower volume
fraction $\phi$ tend to have higher polydispersity $\pol$. This
conclusion is intuitively appealing since higher compression should
disfavour a polydisperse crystalline packing.

\begin{figure}
\begin{center}
\includegraphics[width=8.5cm]{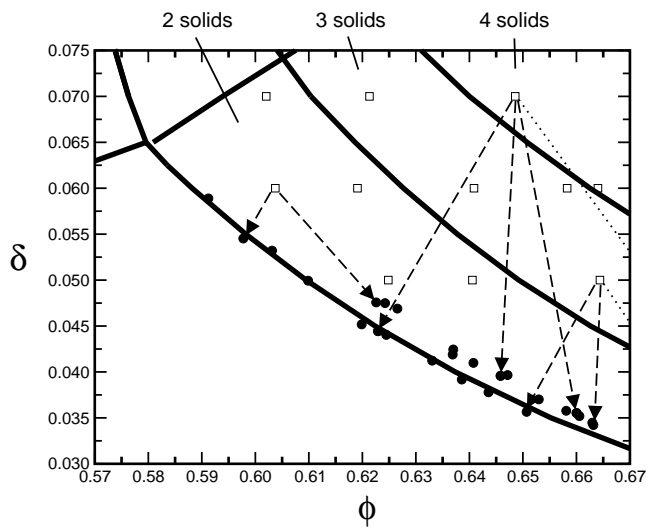}%
\caption{
The properties of the daughter solids (circles) arising
by phase separation from some chosen parents (squares) across the
phase diagram. Plotted are polydispersity $\pol$ versus volume
fraction $\phi$. The arrows show the daughter phases for three parents
explicitly; as indicated by the dotted lines, not all daughter phases
are within the range of the plot. Note the clustering of all daughter
phases near the solid cloud curve.
\label{fig:solid_cluster}}
\end{center}
\end{figure}
We have studied the relation between polydispersity and volume
fraction more quantitatively, by plotting $\pol$ vs $\phi$ for all the
``daughter'' solids that arise by phase separation from a number of
different parents across the phase diagram. We find a set of points
(Fig.~\ref{fig:solid_cluster}) that cluster very closely around the
high-density branch of the solid cloud curve, emphasising the tight
correlation between $\pol$ and $\phi$. Note that some of the points
fall {\em above} the solid cloud curve. This is not a contradiction
because the latter marks the onset of instability against phase
separation only for solids with a triangular size distribution,
whereas the daughter phases plotted here can have rather different
size distributions (compare Fig.~\ref{fig:solid_distribution}).

\begin{figure}
\begin{center}
\includegraphics[width=8.5cm]{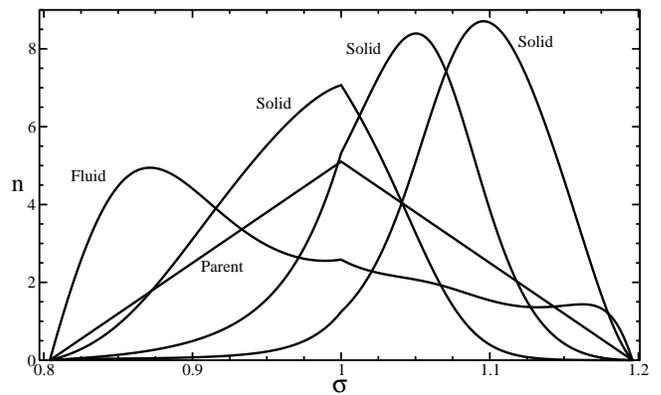}%
\caption{Normalised diameter distributions for F-S-S-S phase
coexistence obtained from a parent distribution with $(\phi, \pol)=(0.603,
0.08)$. From~\protect\cite{FasSol03}.
\label{fig:distribution}
}
\end{center}
\end{figure}
As part of our qualitative overview of fractionation behaviour, we
show next in Fig.~\ref{fig:distribution} the size distributions for a
situation where a fluid coexists with three solids. The general trend
which we observed from the cloud and shadow curves, namely for the
solid(s) to contain the larger particles, is found confirmed
here. However, the details of the fractionation are again nontrivial:
while the coexisting fluid is enriched in the smaller particles as
expected, it also contains ``left over'' large spheres that did not
fit comfortably into the solid phases. It thus in fact ends up having
a {\em larger} polydispersity (0.104) than the parent (0.08) in this
example.

\begin{figure}
\begin{center}
	\includegraphics[width=8cm]{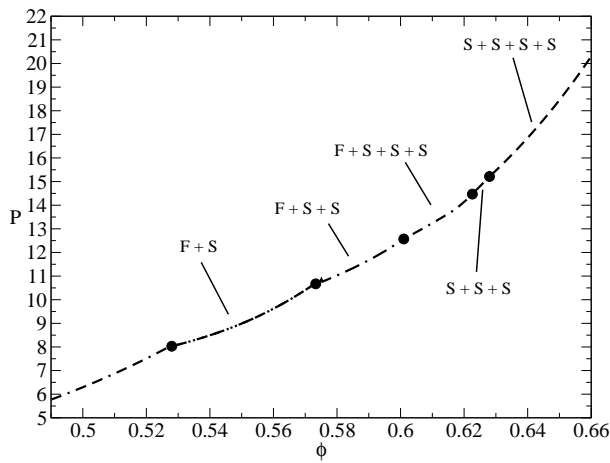}%
	\caption{The osmotic pressure plotted as a function of the
	parent volume fraction along a dilution line, for a triangular
	parent with polydispersity $\pol=0.08$. Phase boundaries are
	marked by full circles; line segments are annotated with the nature
	of the phases (fluid/solid) in the different coexistence
	regions.
\label{fig:pvsphi}
}
\end{center}
\end{figure}
%
Finally, an indirect manifestation of fractionation is provided by the
variation of the osmotic pressure along a dilution line. In a
monodisperse system, the pressure remains constant throughout any
phase coexistence region because the properties of the coexisting
phases do not change; only the fractions of system volume vary which these
phases occupy. In a polydisperse system, on the other hand, the
composition of the coexisting phases varies as the coexistence region
is traversed. We illustrate this in Fig.~\ref{fig:pvsphi} for a
triangular parent size distribution with $\pol=0.08$. It is striking
that the variation of the pressure with volume fraction is almost
smooth, even though a number of phase boundaries are crossed.

\section{Fractionation behaviour \label{sec:3hsFractionationbehaviour}}

We proceed in this section to a systematic study of the fractionation
behaviour of polydisperse hard spheres, having discussed its
qualitative features above. To this end we extend the classical visual
representations in terms of cloud and shadow curves and overall phase
diagrams to include more detailed information about the properties of
the coexisting daughter phases. To obtain insights into the effects of
varying both the parent's volume fraction and its polydispersity, 3-D
plots will be particularly useful here. In a second part we ask
whether there is an optimal way of making the separation between
fractionated phases visible, and suggest principal component analysis
(PCA) as a method for achieving this. We focus throughout on the range
of parent polydispersities $0.04<\pol<0.08$, which covers all the
various coexistence regions in the phase diagram of
Fig.~\ref{fig:parent}.  Where it is necessary to distinguish the
volume fraction and polydispersity of the parent from those of the
daughter phases, we will add the superscript $(0)$, writing
$\phi^{(0)}$ and $\pol^{(0)}$.

\begin{figure}
\begin{minipage}{.49\textwidth}
    \begin{center}  
	\includegraphics[width=8cm]{./phivsphi.eps}%
    \end{center}
\end{minipage}

\vspace{1pt}

\begin{minipage}{.49\textwidth}
  \begin{center}  
      \includegraphics[width=8cm]{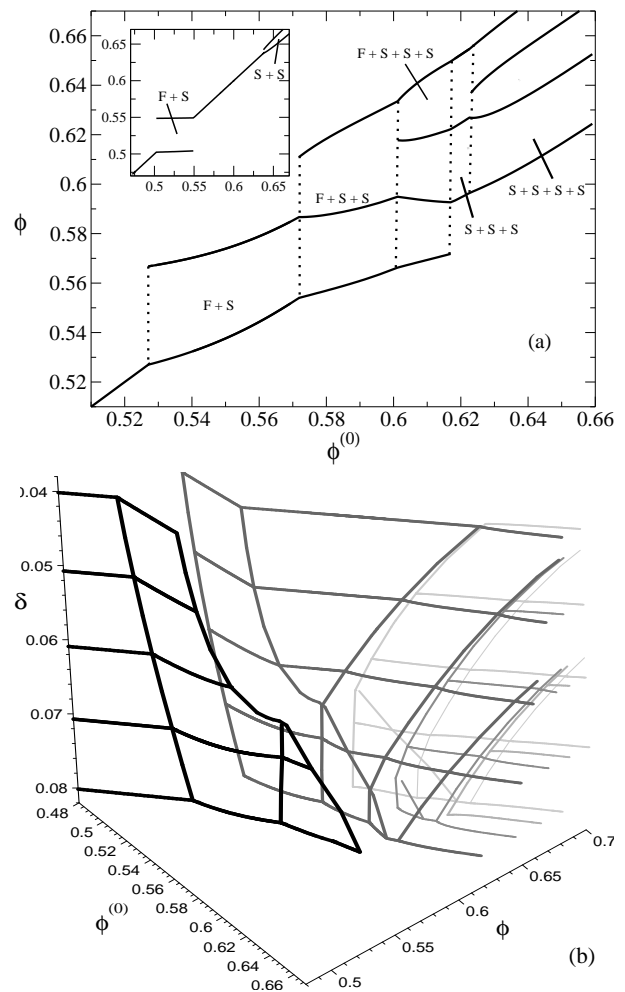}%
  \end{center}  
\end{minipage}
	\caption{(a) Volume fraction $\phi$ of the various coexisting
	daughter phases versus the volume fraction $\phi^{(0)}$ of the
	parent phase, for $\pol = 0.08$ (main graph) and $\pol = 0.04$
	(inset). (b) 3-D plot showing the dependence of the $\phi$
	on $\phi^{(0)}$ and the parent's polydispersity
	$\pol$. Different phases are represented by different grey
	levels. Note that the $\pol$-axis is plotted upside down for
	better visibility. The top and bottom slices correspond to the
	2-D plots in (a).
\label{fig:phivsphi}
}
\end{figure}
%
We start with a 2-D plot showing the volume fraction of the coexisting
phases versus the volume fraction of the parent phase,
Fig.~\ref{fig:phivsphi} (a), for two parent polydispersities
$\pol$. For a narrow parent size distribution ($\pol=0.04$, inset), we
see that the behaviour in the F-S coexistence region is similar to
what would be expected for a monodisperse system, with the volume
fractions of the daughter phases remaining essentially constant. Only
at large $\phi^{(0)}$ does the polydisperse nature of the system
become fully apparent, through the occurrence of S-S phase separation.
For a parent with $\pol=0.08$, on the other hand, the properties of
the daughter phases vary strongly with $\phi^{(0)}$. In the S+S+S and
S+S+S+S regions in particular, the volume fractions of the daughter
solids increase systematically with $\phi^{(0)}$: fractionation from a
denser parent here produces denser daughter phases, rather than
varying proportions of daughters with fixed densities.

To demonstrate more explicitly the change in behaviour as the parent
polydispersity $\pol$ increases, we show in Fig.~\ref{fig:phivsphi}
(b) a 3-D plot of the daughter volume fractions $\phi$ versus
$\phi^{(0)}$ and $\pol$. The orientation of the axes has been chosen
such that horizontal cuts through the plot represent fixed $\pol$,
with the top and bottom planes corresponding to the data shown in the
2-D plots of Fig.~\ref{fig:phivsphi} (a).  A benefit of the 3-D
representation is that each daughter phase now corresponds simply to a
separate surface. Each surface ends at the phase boundary where the
relevant phase disappears from the phase split. The disappearance or
appearance of any phase then causes kinks in the other surfaces. As
expected, only the fluid surface extends to the lowest
$\phi^{(0)}$. As $\phi^{(0)}$ is increased, the ``conventional'' solid
which also exists in the monodisperse limit makes its first
appearance. A further three fractionated solids then eventually appear
one after the other. These are polydispersity-induced, i.e.\ have no
analogue in the monodisperse system, and the surfaces representing
them do not extend to $\pol\to 0$.
%

\begin{figure}
\begin{center}
	\includegraphics[width=8cm]{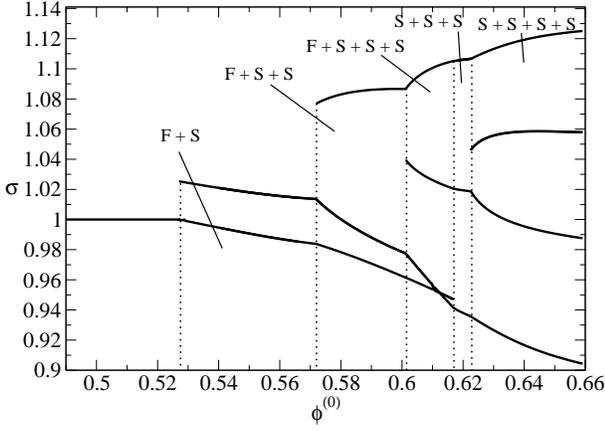}%
	\caption{Mean diameter of coexisting phases plotted against
	parent volume fraction $\phi^{(0)}$, for parent polydispersity
	$\pol=0.08$.  The dashed lines delineate the various phase
	coexistence regions.
\label{fig:rvsphi} }
\end{center}
\end{figure}
%
Having clarified the variation of the volume fractions of the daughter
phases across the phase diagram, we show their mean diameters in
Fig.~\ref{fig:rvsphi}, plotted against parent volume fraction at fixed
(parent) polydispersity $\pol=0.08$. One observes clearly the general
trend for the solid phases to contain larger particles than the
fluid. An exception to this occurs in the F+S+S+S coexistence region,
where the fluid has a slightly larger mean diameter than one of the
solids. The explanation for this can be found in our earlier
discussion of Fig.~\ref{fig:distribution}: in addition to the smallest
spheres, the fluid can also contain some of the larger spheres that
are not accommodated in any of the solids, and this pushes up its mean
diameter. The second qualitative trend demonstrated by
Fig.~\ref{fig:rvsphi} is that the coexisting solids tend to split the
range of particle diameters in the parent distribution amongst
themselves, with almost equidistant mean diameters. As the parent
volume fraction increases, the strength of this fractionation effect
is seen to grow, and the mean diameters become increasingly separated
from each other.

\begin{figure}
\begin{minipage}{.45\textwidth}
    \begin{center} 
	 \includegraphics[width=8cm]{./deltavsphi.eps}	 
    \end{center}
\end{minipage}
\hfill
\begin{minipage}{.45\textwidth}
    \begin{center} 
	 \includegraphics[width=8cm]{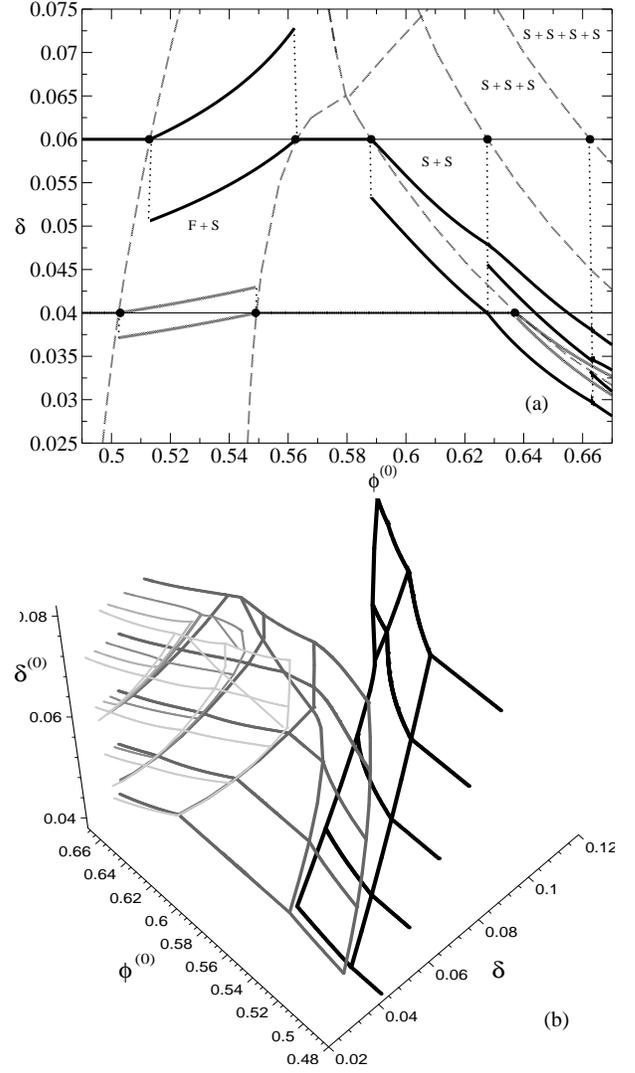}	 
    \end{center}
\end{minipage}
\caption{(a) Plot of the polydispersities of coexisting
	phases along two dilution lines (thin horizontal lines), i.e.\
	as a function of the parent volume fraction for fixed parent
	polydispersity $\pol^{(0)}=0.04$ and $0.06$. The dashed lines
	indicate the phase boundaries in the
	$(\phi^{(0)},\pol^{(0)})$-plane; phases appear or disappear at
	the points where the horizontal line corresponding to the
	fixed parent polydispersity intersects these phase boundaries
	(full circles). (b) Corresponding 3-D plot, showing the
	daughter polydispersities $\pol$ against the parent volume
	fraction $\phi^{(0)}$ and parent polydispersity $\pol^{(0)}$.
\label{fig:deltavsphidelta} }
\end{figure}
Finally, we examine the relationship between the polydispersities
$\pol$ of the different daughter phases and the volume fraction
$\phi^{(0)}$ and polydispersity $\pol^{(0)}$ of the parent.
Fig.~\ref{fig:deltavsphidelta} (a) shows 2-D plots of the daughter
polydispersities versus $\phi^{(0)}$, at $\pol^{(0)}=0.04$ and
$0.06$. As expected, for the more polydisperse parent there are
significant variations of the daughter polydispersities across the
coexistence regions. Where multiple solids coexist, their
polydispersities decrease with increasing parent volume fraction. This
is consistent with the general trend that denser solids tend to be
less polydisperse. In the 3-D plot of Fig.~\ref{fig:deltavsphidelta}
(right), this same trend also causes the surfaces corresponding to the
various solids to have rather similar shapes in the region of
solid-solid coexistence. For the fluid, on the other hand, the graph
demonstrates that it always has a larger polydispersity than the
parent, arising from the presence of large particles ``left over''
from the solid phases (see Fig.~\ref{fig:distribution}).

\subsection{Principal components analysis}

\newcommand{\pr}{r}

The above plots of aspects of fractionation behaviour lead naturally
to the question of whether there is a ``maximally fractionating''
property, i.e.\ one which most strongly reveals the differences
between the various coexisting phases across the phase diagram. We
focus on properties which are generalised moments of the density
distribution, of the form $\pr=\int\!d\sig\,f(\sig)\rhsig$ with some
weight function $f(\sig)$. While not all properties can be expressed
in this way -- the polydispersity $\pol$, for example, involves
squares and ratios of 
such moments -- this is still a fairly large class of measurable
properties; e.g.\ setting $f(\sig)=1$ would give us the number
density, $f(\sig)=\sig^3$ the volume fraction, $f(\sig)=\sig$ the mean
diameter times the number density etc.

Suppose now that we have a number of measurements of $\rhsig$,
specifically the daughter density distributions that arise
within some region of the phase diagram. We can think of the $\rhsig$
as points in a high-dimensional (in fact infinite-dimensional) space,
and of our desired moment $\pr$ as a projection along the direction defined
by $f(\sig)$~\cite{SolWarCat01}. A good choice for a maximally
fractionating property would then be to maximise the {\em variance} of
our moment among the various measured $\rhsig$.  This can be done by
Principal Component Analysis (PCA), a method designed to select
directions of large variance~\cite{Bishop95}. Mathematically, the
requirement of maximum variance can be written as
\begin{equation}
{\rm max}_{f(\sig)} \int\! d\sig\,d\sig'\, f(\sig) A(\sig,\sig')
f(\sig')
\label{eq:max}
\end{equation}
subject to
$
\int\! d\sig\, f^2(\sig)=1;
$
here $A(\sig,\sig')$ is the (infinite-dimensional) covariance matrix 
of our measurements. We define this as
$A(\sig,\sig ') = \langle [\rhsig - \rhzsig ][\rh(\sig ') -
\rh^{(0)}(\sig')]\rangle$. The average here is over all our measurements
of $\rhsig$, and we subtract off for each $\rhsig$ the
corresponding parent distribution $\rhzsig$. This effectively removes
the average of the various measured $\rhsig$ because, from particle
conservation\eq{particle_cons}, the parent is a weighted average over
the various daughter phases. An alternative definition of $A$ would be
to remove the actual measurement average, $A(\sig,\sig ') = \langle
[\rhsig - \langle\rhsig\rangle][\rh(\sig ') -
\langle\rh(\sig')\rangle]\rangle$. In our numerical experiments
described below, this lead to almost indistinguishable results.

The maximisation problem\eq{eq:max} is in principle over an
infinite-dimensional function space. To arrive at a more practical
task, we restrict the search to a subspace by requiring the weight
function to be of the form $f(\sig) = \sum_{i=0}^3
\alpha_i\sig^i$. This corresponds to searching for a maximally
fractionating property among those expressible as linear combinations
of the moments $\rh_0, \ldots, \rh_3$, i.e.\ $\pr = \sum_{i=0}^3
\alpha_i \rh_i$  With this simplification, the problem\eq{eq:max}
reduces to 
\begin{equation}
{\rm max}_{\alpha}\ \alpha^T C \alpha 
\quad \mbox{subject to} \quad \alpha^T D \alpha=1
\label{eq:low_dim_PCA}
\end{equation}
Here $\alpha$ denotes the vector with elements $\alpha_0, \ldots,
\alpha_3$ and the $4 \times 4$ matrix $C$ is defined as
\begin{eqnarray*}
C_{ij}=
\int\! d\sig \,d\sig'\,\sig^i A(\sig,\sig') (\sig')^j
= \langle(\mi - \rh^{(0)}_i)(\rh_j - \rh^{(0)}_j)\rangle
\end{eqnarray*}
which is just the covariance matrix of the moments, with the parent
moments again subtracted off. The matrix $D$, on the other hand, is
given by $D_{ij}=\int\! d\sig\, \sig^{i+j}$. The $\sig$-integration
range has to be bounded to make this well-defined. In our case of a
triangular parent distribution the obvious choice, adopted here, is to
make this range equal to the range of particle sizes occurring in the
parent.

Imposing the constraint in\eq{eq:low_dim_PCA} via a Lagrange
multiplier shows that solution vectors $\alpha$ must obey $C\alpha =
\lambda D\alpha$, or equivalently $D^{-1/2}CD^{-1/2}(D^{1/2}\alpha) =
\lambda D^{1/2}\alpha$. The solutions can thus be obtained by an
eigenvalue decomposition of the matrix $D^{-1/2}CD^{-1/2}$, with
$\lambda$ the eigenvalue and $D^{1/2}\alpha$ the corresponding
eigenvector. (Numerically, it is more convenient to solve the
equivalent problem of finding the eigenvalues and right eigenvectors
of the matrix $D^{-1}C$.)
The eigenvectors are termed principal components, and the $\lambda$'s
give the variance captured by each principal component. The most
important principal component, and the one of interest to us, is then
the one with the largest $\lambda$.

%
\begin{figure}
\begin{center}
\includegraphics[width=8.5cm]{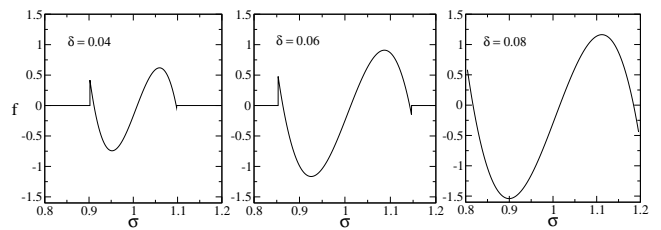}
\caption{Principal component weight function, $f(\sig)$,
	obtained from data along dilution lines for three different
	values of parent polydispersity 0.04, 0.06 and 0.08. The range
	of $\sig$-values where $f(\sig)\neq 0$ is in each case that
	over which the parent density distribution is nonzero, i.e.\
	the range of particle sizes actually occurring in the system.
\label{fig:pcafunction}
}
\end{center}
\end{figure}
%
We have implemented this PCA search for maximally fractionating
properties by considering as our measured $\rhsig$ the daughter
phases as they occur along a dilution
line. We do this separately for triangular parent distributions of
polydispersity 0.04, 0.06 and 0.08, respectively, because different
ranges of particle size $\sig$ are relevant in the three cases. The
resulting weight functions $f(\sig)$ are plotted in
Fig.~\ref{fig:pcafunction}. One sees that in all cases, $f(\sig)$ is
to a good approximation a combination of the odd weight functions
$\sig$ and $\sig^3$, with the coefficients such that $f(\sig)$ crosses
zero near the edge of the $\sig$-range. Loosely speaking, the function
$f(\sig)$ can be interpreted as an approximation to ${\rm
sgn}(\sig-1)$ within the space spanned by $\sig^0, \ldots, \sig^3$,
i.e.\ by a third-order polynomial in $\sig$. It thus effectively
measures the difference in number density between particles above and
below the mean parental diameter. This is an intuitively appealing
measure of fractionation behaviour.

\begin{figure}
\begin{center}
	\includegraphics[width=8.8cm]{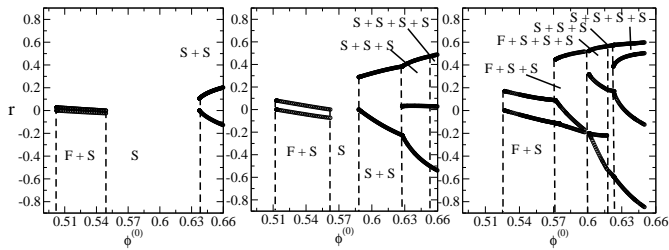}%
	\caption{Maximally fractionating moment (as selected by PCA)
	for coexisting daughter phases, plotted against parent volume fraction
	$\phi^{(0)}$ at parent polydispersity $\pol=0.04$, 0.06 and
	0.08 (from left to right).  
 \label{fig:pcaprojection}}
\end{center}
\end{figure}

Finally, Fig.~\ref{fig:pcaprojection} shows the properties of the
daughter phases as measured by the maximally fractionating observable
selected by PCA. The overall features of the plot on the right, for
parent polydispersity $\pol=0.08$, are not dissimilar to the mean
diameter representation in Fig.~\ref{fig:rvsphi}, so that the benefit
of PCA in this problem is relatively modest. Some interesting features
are accentuated by PCA, however; e.g.\ the crossover between the fluid
and solid lines is more pronounced in Fig.~\ref{fig:pcaprojection},
demonstrating clearly how the fluid size distribution acquires a
significant fraction of the larger particles. We expect that the
benefits the PCA method of selecting maximally fractionating properties 
should become more pronounced in systems with several polydisperse
attributes $\sig$, e.g.\ particle size and charge. Suitable properties
for revealing fractionation behaviour could then depend on
combinations of these attributes, which can be systematically found
using PCA.

\section{Comparison with perturbative theories and Monte Carlo simulations}
\label{sec:comparison}

In this section we validate our theoretical predictions in two
ways. First, we compare to perturbative theories for near-monodisperse
parents, which predict qualitatively how the properties of coexisting
phases should vary as the parent is made increasingly
polydisperse. Second, we compare quantitatively to Monte Carlo
simulations of polydisperse hard spheres with an imposed chemical
potential distribution.

\subsection{Near-monodisperse systems}

For systems that are nearly monodisperse, one can make general
statements about the fractionation behaviour by considering deviations
of particle diameters from the mean as small and performing a
perturbation expansion~\cite{EvaFaiPoo98,Evans01}. This approach
presupposes, of course, that the phase separation of interest occurs
already in the monodisperse reference system. In our hard sphere case it is
therefore applicable only to fluid-solid coexistence. Phase separation
involving several solid phases is induced by polydispersity itself and
cannot be treated perturbatively.

A strong prediction of the perturbative approach is that the
difference in mean particle diameters of two coexisting phases,
$\Delta\bar{\sig}=\bar{\sig}^{(1)}-\bar{\sig}^{(2)}$ is {\em universally}
related to the parental polydispersity $\pol$ via
\begin{equation}
\Delta\bar{\sig} \propto \pol^2.
\label{eq:powlaw2}
\end{equation}
An increase in the width of the parent size distribution thus
contributes only at second order to the mean diameter difference. The
proportionality coefficient in this relation is non-universal and
depends on the properties of the phase separation in the corresponding
monodisperse reference system. Equation~(\ref{eq:powlaw2}) is the
leading term in a perturbation expansion in $\pol$ at fixed {\em
density} of the parent. The {\em volume fraction} $\phi$ then varies
with $\pol$;
for symmetric size distributions it increases. If instead $\phi$ is
held constant as $\pol$ is varied, the perturbation expansion is
modified, though the leading term (\ref{eq:powlaw2}) remains
unaffected.

A relation analogous to equation~(\ref{eq:powlaw2}) applies to the difference
in polydispersities $\Delta\pol = \pol^{(1)}-\pol^{(2)}$ of the
daughter phases. For symmetric parent size distributions, one finds
\begin{equation}
\Delta{\pol} \propto \pol^3\;.
\label{eq:powlaw3}
\end{equation}
so that an increase in the parent polydispersity $\pol$ only affects
$\Delta\pol$ at {\em third} order.

\begin{figure}
\begin{center}
\includegraphics[width=8.5cm]{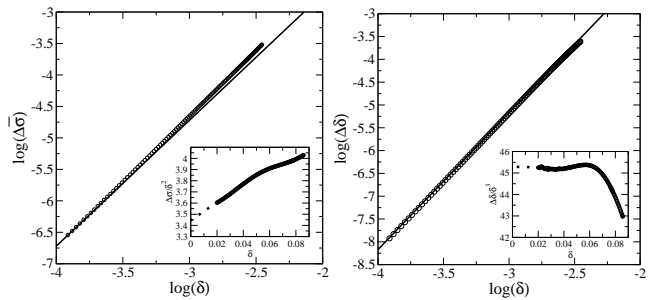}
\end{center}
\caption{Log-log plots of the difference in mean diameters
(left) and in polydispersity (right) for coexisting fluid-solid phases
at parent density $\rh=0.52$, plotted against parent polydispersity
$\pol$. The solid lines show the theoretically expected power laws,
Eqs.~(\ref{eq:powlaw2}) and (\ref{eq:powlaw3}), with the proportionality
constants fitted by eye to the numerical data.  The insets show the
ratios $\Delta\bar{\sig}/\pol^2$, $\Delta{\pol}/\pol^3$, which
from the theory are expected to approach constants for $\pol\to
0$. The data are consistent with this (note the narrow ranges
displayed on the $y$-axes). 
\label{fig:powlaw}
}
\end{figure}
To verify the above perturbative predictions, we determined the mean
diameters and polydispersities of the daughter phases for fluid-solid
separation at parent density $\rh=0.52$ and parent polydispersity
$\pol$ ranging from $0.02$ to $0.086$.  Fig.~\ref{fig:powlaw} shows
log-log plots of $\Delta\bar{\sig}$ and $\Delta\pol$ against $\pol$,
confirming the predicted power laws.  The insets show the ratios
$\Delta\bar{\sig}/\pol^2$ and $\Delta\pol/\pol^3$. Our data are
again consistent with the theoretical expectation that these ratios should
approach constants in the limit of small $\pol$. Overall, the 
scaling predictions of perturbative theories for near-monodisperse
systems are fully obeyed by our data.

\begin{figure}
  \begin{center}
  \includegraphics[width=8cm]{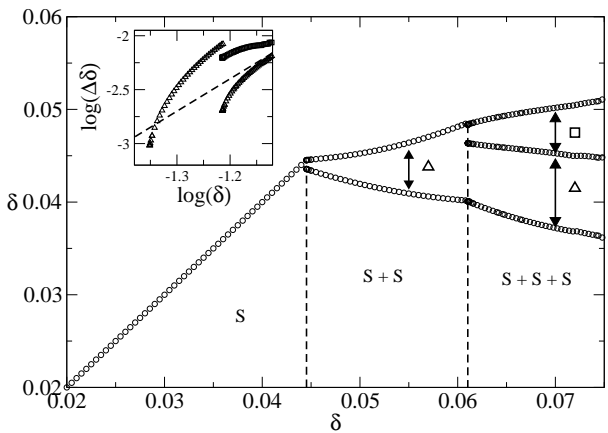}
  \end{center} \caption{Plot of the polydispersity of coexisting
  solids versus the polydispersity $\pol$ of the parent phase. The
  parent number density is kept constant at $\rho=0.62$ while the
  latter is varied. The inset shows the log-log plot of the difference
  of the polydispersities $\Delta\pol$ in the S+S and S+S+S region
  versus $\pol$ and the slope of the theoretically expected power
  law\eq{eq:powlaw3} (dashed line). This demonstrates that the
  perturbative prediction for the scaling of $\Delta\pol$ with $\pol$
  does not apply here. \label{fig:deltavsdelta}}
\end{figure}

Finally, to emphasise the point that perturbative approaches do not
apply to phase separations that are caused by polydispersity, we show
in Fig.~\ref{fig:deltavsdelta} the evolution of the polydispersity of
coexisting solids as the parent polydispersity is increased, at fixed parent
density $\rh=0.62$. As the inset shows, there is now no simple
relationship akin to\eq{eq:powlaw3} which relates the difference in
the polydispersities of the daughter solids to the $\pol$ of the
parent. In fact, the perturbative limit of small $\pol$ is not even
defined here, since the solid-solid phase separation only occurs above
a nonzero (density-dependent) threshold value of $\pol$.

\subsection{Comparison with Monte Carlo simulations}
\label{sec:3hsComparisonwithBolhuisandKofkeresults}

As discussed in the preceding sections, our theoretical results for
fluid-solid coexistence in polydisperse hard sphere systems are in
qualitative agreement with numerical
simulations~\cite{BolKof96,BolKof96b,KofBol99}, in particular
concerning the coexistence of rather polydisperse fluids with solids
that have a much narrower size distributions. There is, however, an
important difference: our calculations apply to the experimentally
realistic case where an overall parent density distribution is
fixed. The simulations, on the other hand, are performed at imposed
chemical potential differences, with the actual size distributions in
the coexisting phases varying strongly across the phase diagram. In
order to obtain a quantitative comparison between theory and
simulations, we calculate explicitly in this section the theoretical
predictions for the -- somewhat unrealistic -- scenario addressed in
the simulations. We will find good quantitative agreement, thus
validating our approach and, in particular, our choice of free energy
expressions for the fluid and solid phases.

The simulations of~\cite{BolKof96,BolKof96b,KofBol99} were carried out
in an isobaric semi-grandcanonical ensemble, which corresponds to
fixed particle number $N$, pressure $P$ and chemical potential
differences $\mu(\sig)-\mu(\sig_b)$. Here $\sig_b$ is the diameter of
a reference particle. The advantage of the semi-grandcanonical
ensemble is that it allows many different realizations of the particle
size distribution to be sampled, thus minimising finite-size
effects. The fixed particle number $N$, on the other hand, avoids
simulation moves where particles need to be inserted into dense fluids
or solids.

Bolhuis and Kofke~\cite{BolKof96,BolKof96b} considered specifically a
quadratic form for the chemical potential differences,
\begin{equation}
\mu(\sig)-\mu(\sig_b) = - \frac{(\sig - \sig_b)^2}{2 \nu}.
\label{eq:bolkofchepot}
\end{equation}
The activity $\exp[\mu(\sig)]$ thus has a Gaussian shape of variance
$\nu$. For small $\nu$, one expects the activity distribution to set
the size distributions in the coexisting phases, which should
therefore have polydispersity $\pol=\nu^{1/2}$; $\nu\to 0$ recovers
the monodisperse case. The reference diameter $\sig_b=1$ is held fixed
as $\nu$ is increased from zero. The pressure $P$ is then adapted by
Gibbs-Duhem integration~\cite{Kofke93} to follow the line of
fluid-solid phase coexistence in the $(\nu,P)$-plane.

In order to reproduce the situation considered in the
simulations using our theoretical approach, we will study a
system with prior $\prior = \exp[-(\sig-1)^2/(2\nu)]$. The moment
free energy then gives the free energy of phases with density
distributions of the form (cf.\eq{family})
\begin{displaymath}
\rhsig = 
\exp\left[-\frac{(\sig-1)^2}{2\nu} + \sum_{i=0}^3 \lambda_i
\sig^i \right]
\end{displaymath}
From\eq{musig}, the corresponding chemical potentials have the form
\begin{equation}
\mu(\sig) = \ln \rhsig + \sum_{i=0}^3 \muex_i \sig^i
= -\frac{(\sig-1)^2}{2\nu} + \sum_{i=0}^3 (\lambda_i+\muex_i)\sig^i
\label{BolKofmu}
\end{equation}
Now the $\lambda_i+\muex_i$ are just the moment chemical potentials
$\mu_i=\partial\fmom/\partial\mi$. So if we apply the moment free
energy but treat the moment densities $\rh_1$, $\rh_2$, $\rh_3$ as
non-conserved, the associated $\mu_i$ are forced to vanish
automatically at equilibrium. The sum over $i$ in\eq{BolKofmu} then
reduces to a constant, 
$\mu_0=\lambda_0+\muex_0$, and we have precisely the chemical
potential differences\eq{eq:bolkofchepot} used in the simulations. In
summary, applying the MFE method with a Gaussian prior and $\rh_0$ the
only conserved moment, we increase $\rh_0$ from zero until coexistence
with a solid phase is first found. This is then the desired
fluid-solid coexistence for quadratic chemical potential differences,
and we can determine in particular the pressure $P$ at coexistence. 
Repeating this process for a range of values of $\nu$ gives the
coexistence curve in the $(\nu,P)$-plane.

Our actual implementation of this approach has one minor
difference. In the simulations it is observed that the mean diameters
in the coexisting phases decrease significantly as $\nu$ is increased,
eventually becoming much smaller than $\sig_b$. For our numerical
work, however, it is desirable to keep the size distributions within a
fixed range, e.g.\ in order to ensure that our chemical potentials for
the solid phase remain reliable. To achieve this, we treat not just
$\rh_0$ but also $\rh_1$ as conserved. Keeping $\rh_1/\rh_0=1$ as
$\rh_0$ is varied then ensures that the fluid phase always has unit
mean diameter, and the particle sizes in the coexisting solid are
expected to be comparable. This ensures that we can use a fixed
$\sig$-range for all calculations, for which we choose $\sig\in
[0.7,1.3]$.

With $\rh_0$ and $\rh_1$ both conserved, the chemical
potentials\eq{BolKofmu} become
\begin{eqnarray}
\mu(\sig) & = & -\frac{(\sig-1)^2}{2\nu} + \mu_0 + \mu_1\sig \\
 & = & -\frac{(\sig-1-\nu\mu_1)^2}{2\nu} + \frac{1}{2} \nu\mu_1^2 + \mu_1  + \mu_0
\end{eqnarray}
which is again of the form\eq{eq:bolkofchepot} but now with a varying
reference diameter $\sig_b = 1+\nu\mu_1$. The corresponding scaled
quantities that are to be compared to $\nu$ and $P$ from the simulations
are then $\nu/\sig_b^2$ and $P\sig_b^3$
\footnote{%
In fact, because of our inclusion of a factor $\pi/6$ in the unit
volume $\univ = (\pi/6)\unil^3$, the simulation values of $P$ have to
be multiplied by $\pi/6$ for comparison with our results; this is what
we do below.}%
. Note finally that our numerical implementation again uses
centred moments, with weight functions $[(\sig-1)/\pol_0]^i$ rather
than $\sig^i$, but this causes no conceptual differences. In
particular, keeping the standard moments $\rh_0$ and $\rh_1$ conserved
is equivalent to conservation of the centred moments with $i=0$ and
$i=1$, because of the linear relations between the two sets of
moments.

\begin{figure}
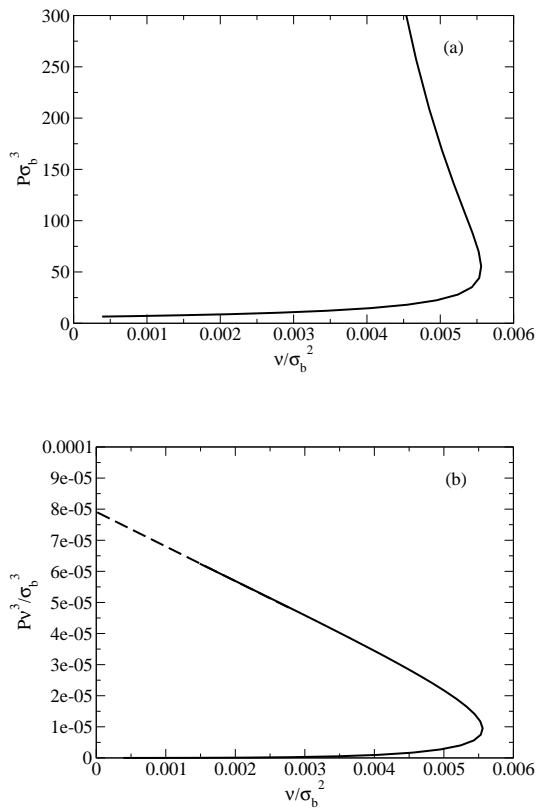

\begin{minipage}{.45\textwidth}
    \begin{center} 
	 \includegraphics[width=7cm]{./p_vs_nu.eps}	 
    \end{center}
\end{minipage}

\vspace{24pt}

\begin{minipage}{.45\textwidth}
  \begin{center} 
    \includegraphics[width=7cm]{./p_vs_nu_2.eps}	
  \end{center}  
\end{minipage}
\caption{(a) Solid-fluid coexistence pressure $P\sig_b^3$ as a function
of the imposed width $\nu/\sig_b^2$ of the activity distribution. Both
are scaled appropriately with $\sig_b$ to account for the fact that
$\sig_b$ varies in our calculation but is held constant in the
simulations. (b) The pressure is rescaled to
$P\sig_b^3(\nu/\sig_b^2)^3 = P
\nu^3/ \sig_b^3$ to show the limiting behaviour for $\nu/\sig_b^2\to 0$.
\label{fig:pvsnu}
}
\end{figure}

Fig.~\ref{fig:pvsnu}~(a) shows our results for the coexistence curve in
the $(\nu/\sig_b^2,P\sig_b^3)$-plane. As $\nu$ increases (starting
from the bottom left corner), both $P\sig_b^2$ and $\nu/\sig_b^2$
initially increase. However, eventually $\nu/\sig_b^2$ reaches a
maximum value $\nu_{\rm max}=\nu/\sig_b^2=0.0056$. At this point, the
slope $d(P\sig_b^3)/d(\nu/\sig_b^2)$ becomes infinite.  On further
increasing $\nu$, the coexistence curve then bends back, with
$\nu/\sig_b^2$ decreasing towards zero while the pressure diverges.
Bolhuis and Kofke~\cite{BolKof96} argued that this divergence arises
because the pressure is measured on the scale of the mean $\sig_b$ of
the activity distribution, while the typical particle diameters in the
coexisting phases become much smaller than $\sig_b$, by a factor
scaling as $\nu/\sig_b^2$. The rescaled pressure
$P\sig_b^3(\nu/\sig_b^2)^3 = P\nu^3/ \sig_b^3$ should therefore
approach a constant value in the limit $\nu/\sig_b^2\to 0$. The
simulations were consistent with this expectation, and our theoretical
results in Fig.~\ref{fig:pvsnu} (b) are in full agreement. By
extrapolation, we estimate the limiting or `terminal' value of
the rescaled pressure, i.e.\ the point where the rescaled coexistence
curve intersects the vertical axis, as $P_{\rm t}=7.9\times10^{-5}$.
%
%

\begin{figure}
    \begin{center}
    \includegraphics[width=8.5cm]{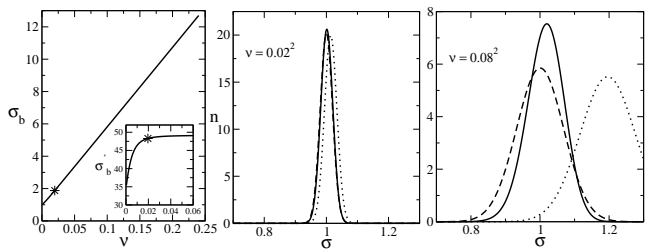}
    \caption{Left: Plot of $\sig_b$ as a function on $\nu$. The inset
    shows the derivative $d\sig_b/d\nu$. As expected, $\sig_b$ grows
    linearly as $\nu$ becomes large. The star indicates the value of
    $\nu$ at which $\nu/\sig_b^2$ reaches its maximum and the slope of
    the pressure plots in Fig.~\protect\ref{fig:pvsnu} becomes
    infinite.  Middle and right: Normalised size distributions $\nsig$
    of the coexisting fluid (dashed line) and solid (solid line)
    phases. The dotted curve gives the shape of the activity
    distribution $\exp(\musig)$. Middle: For $\nu=0.02^2$, fluid and
    solid have essentially identical size distributions of
    polydispersity $\pol\approx\nu^{1/2}=0.02$; the corresponding
    value of $\sig_b$ is 1.02. Right: For $\nu=0.08^2$, the size
    distributions are significantly different from each other and from
    the activity distribution, which is now centred around
    $\sig_b=1.19$.
\label{fig:bolkofdistribution}
}  
\end{center}
\end{figure}
The scaling mentioned above implies that, as $\nu$ becomes large, the
mean particle diameter in the coexisting phases will be of order
$\sig_b(\nu/\sig_b^2) = \nu/\sig_b$, rather than $\sig_b$. In our
scheme, where the mean diameter in the fluid is fixed at unity,
$\sig_b$ should thus become linear in $\nu$. As shown in
Fig.~\ref{fig:bolkofdistribution} (left), this is indeed what we
find. A plot of the numerical derivative of this dependence, in the
inset of Fig.~\ref{fig:bolkofdistribution} (left), also reveals that at small
$\nu$-values -- below those where $\nu/\sig_b^2$ reaches its maximum
-- the behaviour is no longer exactly linear.
This is to be expected considering that $\sig_b=1+\nu\mu_1$ depends on
both $\nu$ and $\mu_1$.

The plots in the middle and on the right of
Fig.~\ref{fig:bolkofdistribution} show particle size distributions in
the coexisting fluid and solid phases at two different values of
$\nu$. For small $\nu=0.02^2$, the distributions are essentially
identical and have width $\pol\approx\nu^{1/2}=0.02$ as expected; they
are also close to the activity distribution, which has its peak at
$\sig_b=1.02$ for this $\nu$. For larger $\nu=0.08^2$, on the other
hand, there is significant fractionation between the fluid and the
solid. One can now also clearly see how the mean particle diameters --
which are exactly unity in the fluid, by construction, and around 1.02
in the solid -- become smaller than the mean of the activity
distribution, which is $\sig_b=1.19$ for this value of $\nu$.

\begin{figure}
    \begin{center}
	 \includegraphics[width=7cm]{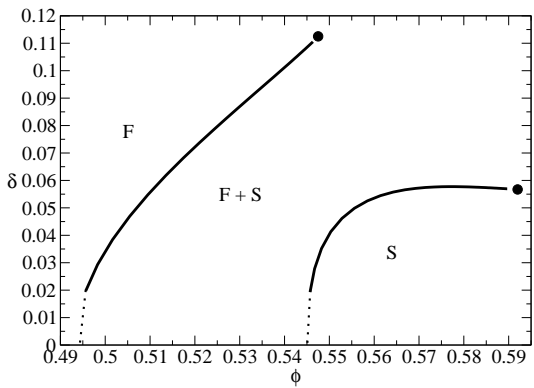}
	 \caption{Phase diagram for fluid-solid coexistence with
imposed quadratic chemical potential differences. Plotted is the
polydispersity $\pol$ versus the volume fraction $\phi$ of the
coexisting fluid and solid phases; $\nu$ increases from bottom left to
top right along the curves. The circles indicate the terminal points
reached by extrapolating to $\nu\to\infty$. The dotted lines sketch
the approach to the known monodisperse limit $\nu\to 0$. 
\label{fig:bolkofphadia}
}  
\end{center}
\end{figure}

To summarise the properties of the coexisting fluid and solid phases,
we plot them in a volume fraction--polydispersity phase diagram, shown
in Fig.~\ref{fig:bolkofphadia}. 
%
As discussed in detail in~\cite{BolKof96}, a feature which is at first
surprising is that the curves terminate, with the properties of the
coexisting phases approaching finite limits for $\nu\to\infty$. One
has to bear in mind, however, that the occurrence of such terminal
points is directly linked to the shape of the imposed chemical
potential distribution and so not physically very meaningful. Indeed,
other shapes can and do give fluids and solids with larger $\phi$ and/or
$\pol$~\cite{KofBol99}. 

We obtain the location of the terminal points by plotting our
numerical predictions against $1/\nu$ and extrapolating to $1/\nu=0$%
. 
\begin{table}
\begin{center}
\begin{tabular}{|c|c|c|}
\hline
Quantity & Bolhuis and Kofke~\cite{BolKof96} & Present work \\
\hline
$\nu_{\rm max}$  &      0.0056       &     0.0056    \\
$P_{\rm t}$    &  7.9 $\times 10^{-5}$    &  7.9 $\times 10^{-5}$ \\
$(\phi_{\rm t,f}, \pol_{\rm t,f})$ & (0.545, 0.12)  &  (0.548, 0.113) \\
$(\phi_{\rm t,s}, \pol_{\rm t,s})$ & (0.575, 0.057) &  (0.592, 0.057)\\
\hline
\end{tabular}
\caption{
Comparison between some characteristic quantities of fluid-solid
coexistence at imposed chemical potential distribution, as determined
in simulations~\cite{BolKof96} and in the present theoretical study.
Here $\nu_{\rm max}$ is the maximum value of $\nu/\sig_b^2$, $P_{\rm
t}$ the terminal value of the rescaled pressure $P\nu^3/\sig^3_b$ in
the limit $\nu/\sig_b^2\to 0$, $\phi_{\rm t,f/s}$ the terminal volume
fraction for the fluid/solid phases and $\pol_{\rm t,f/s}$ the
corresponding terminal polydispersity.
\label{tab:bolkofcomparison}
}
\end{center}
\end{table}
%
The resulting values are compared in Table~\ref{tab:bolkofcomparison}
with those obtained in the simulations of~\cite{BolKof96}. We find
excellent quantitative agreement for $\nu_{\rm max}$, defined as the
maximum value of $\nu/\sig_b^2$, and $P_{\rm t}$, the terminal value
of the rescaled pressure $P\nu^3/\sig^3_b$. Similar comments apply to
the volume fractions and polydispersities at the terminal points of
the fluid and solid coexistence curves in Fig.~\ref{fig:bolkofphadia}.
Only the terminal volume fraction of the solid is over-estimated
somewhat, but even here the deviation is less than $3\%$.

In conclusion, our theoretical predictions for fluid-solid coexistence
at imposed chemical potential differences are not just in qualitative
but in fact quantitative agreement with the outcomes of Monte Carlo
simulations. This provides strong validation for our approach. It
demonstrates in particular that our chosen model free energies for the
hard sphere fluid and solid are accurate, at least in the range of
relatively small polydispersities studied here.

\section{Conclusion and outlook}
\label{sec:conclusion}

We have studied the equilibrium behaviour of size-polydisperse hard
spheres, starting from accurate free energy expressions for the hard
sphere solid and fluid. Cloud and shadow curves, which locate the
onset of phase coexistence, were found exactly by using the
moment free energy (MFE) method. We were also able to calculate the
full phase diagram, however, by using the MFE results as starting
points for a solution of the full phase equilibrium equations.

In contrast to earlier simplified theoretical treatments, we found no
point of equal concentration between fluid and solid. Rather, the
fluid cloud curve continues to larger polydispersities while the
coexisting solid shadow always has a polydispersity $\pol$ below a
``terminal'' value of around $\polt\approx 0.06$. In this sense the
concept of terminal polydispersity only applies to the solid phase,
while any experimentally observed terminal polydispersity from the
fluid side must be attributed to non-equilibrium effects such as an
intervening kinetic glass transition~\cite{PusVan87}, large nucleation
barriers~\cite{AueFre01} or the unusual growth kinetics of
polydisperse crystals~\cite{EvaHol01}.

Concomitant with the absence of the point of equal concentration, we
also found no re-entrant melting. Instead, a sufficiently compressed
polydisperse solid fractionates into two or more solid phases; our
results in this region of the phase diagram are consistent with
previous approximate calculations. In addition, we found that
coexistence of several solids with a fluid phase is also
possible. That such phase splits must exist is clear from the fact
that the solid cloud curve has two branches, describing onset of
fluid-solid and solid-solid phase separation at low and high densities
respectively; a fluid-solid-solid coexistence region begins where
these meet.

We then analysed the fractionation behaviour in detail. As a general
rule, the fluid phases contain the smaller particles in the system,
while the larger ones are found predominantly in the solid phases. The
solid phases have smaller polydispersities $\pol$ than the parent
phase; this is as expected since narrower particle size distributions
are more easily accommodated on a regular lattice. Consistent with
this physical intuition, we also found that there is a strong
correlation between the polydispersity $\pol$ and the volume fraction
$\phi$ of coexisting solids, with the denser phases (larger $\phi$)
having smaller $\delta$. For the fluid phases, on the other hand, we
found {\em larger} polydispersities than in the parent.
This is because the fluid contains, together with a
relatively narrow distribution of smaller particles, also residual
larger particles that were not incorporated into any of the solid
phases.

Three-dimensional fractionation plots transparently showed the
continuity of the properties of the various phases across the phase
diagram, with each corresponding to a distinct surface. The individual
phases change significantly as coexistence regions are traversed; this
is in contrast to monodisperse systems, where only the amounts of
coexisting phases vary. Correspondingly, the pressure in the
polydisperse case was seen to vary almost smoothly on traversing
several coexistence regions, whereas it would be constant within each
for a monodisperse system. We finally proposed a method for
constructing maximally fractionating observables, i.e.\ measurable
properties which reveal most clearly the differences between the
various coexisting phases. This was based on Principal Components
Analysis in the space of the relevant density distributions. The
benefits of this method were modest in our case, but it could be of
significant interest for analysing systematically the phase behaviour
of systems with more than one polydisperse attribute, e.g.\ particle
size and charge.

In the final section we compared our predictions to perturbative
theories for near-monodisperse systems, finding full agreement. We
also performed a detailed comparison with Monte Carlo simulation
carried out at imposed chemical potential distribution, where particle
size distributions vary across the phase diagram. The excellent
agreement obtained provided strong validation of our approach and in
particular of our choice of model free energies for polydisperse hard
sphere fluids and solids.

There are a number of possibilities for extending and complementing
the present work. Our study was limited to systems with relatively
narrow size distributions, with polydispersities $\pol$ up to $\approx
0.14$. At higher $\pol$, fluid-fluid demixing would eventually be
expected to occur~\cite{Warren99,Cuesta99}. So far only the spinodals
for this have been calculated, however, and it would be interesting to
understand the topology of the full phase diagram in this large-$\pol$
region. One might, for example, expect to find coexistence of multiple
fluids, but the conditions required for this are at present unclear.

Quantitative studies of the phase behaviour of hard spheres at large
$\pol$ would require accurate model free energies for wide particle
size distributions. For the fluid, the BMCSL approximation may
continue to be sufficient, although a recent comparison with
simulations has revealed some shortcomings~\cite{WilSol02}. Much more
pressing is the need for an accurate free energy for strongly
polydisperse hard sphere solids. This would allow one to investigate,
for example, whether the dominance of the largest particles at the
onset of solid-solid coexistence which we found for Schultz size
distributions is a genuine physical effect. A quantitative
verification of the prediction that polydisperse hard spheres with
sufficiently fat-tailed size distributions split off multiple
fractionated solids even at low density~\cite{Sear99c} would also be
of interest. A significant challenge in the construction of
approximate free energies for hard sphere solids is that the
simplifying assumption of a substitutionally disordered structure --
which was implicit in our study -- may break down at
large polydispersities. Competing substitutionally ordered structures
would then also have to be considered.

Finally, it will be exciting to generalise our approach to more
complex colloidal systems, by for example including attractive
interactions or extending the scope to polydisperse colloid-polymer
mixtures. Work on these scenarios is currently under way.


We are grateful to Paul Bartlett for making his numerical code for the
solid free energy available to us, and to EPSRC for financial support
(GR/R52121/01).


\bibliography{bibliography}

\begin{thebibliography}{10}

\bibitem{HanMcD86}
J.~P. Hansen and I.~R. McDonald, {\em Theory of Simple Liquids} (Academic
  Press, New York, 1986).

\bibitem{RusSavSch89}
W.~B. Russel, D.~A. Saville, and W.~R. Schowalter, {\em Colloidal dispersions}
  (Cambridge University Press, Cambridge, 1989).

\bibitem{PusVan86}
P.~N. Pusey and W. Van~Megen, Nature {\bf 320},  340  (1986).

\bibitem{PauAck90}
S.~E. Paulin and B.~J. Ackerson, Phys.\ Rev.\ Lett. {\bf 64},  663  (1990).

\bibitem{Goetze91}
W. G{\"{o}}tze,  in {\em Liquids, freezing and glass transition}, edited by
  J.~P. Hansen, D. Levesque, and J. Zinn-Justin (North-Holland, Amsterdam,
  1991), pp.\ 287--503.

\bibitem{Pusey91}
P.~N. Pusey,  in {\em Liquids, freezing and glass transition}, edited by J.~P.
  Hansen, D. Levesque, and J. Zinn-Justin (North-Holland, Amsterdam, 1991).

\bibitem{DicPar85}
E. Dickinson and R. Parker, J.\ Phys.\ Lett. {\bf 46},  L229  (1985).

\bibitem{BolKof96}
P.~G. Bolhuis and D.~A. Kofke, Phys.\ Rev.\ E {\bf 54},  634  (1996).

\bibitem{BolKof96b}
P.~G. Bolhuis and D.~A. Kofke, J.\ Phys.\ Cond.\ Matt. {\bf 8},  9627  (1996).

\bibitem{PhaRusZhuCha98}
S.~E. Phan, W.~B. Russel, J.~X. Zhu, and P.~M. Chaikin, J.\ Chem.\ Phys. {\bf
  108},  9789  (1998).

\bibitem{KofBol99}
D.~A. Kofke and P.~G. Bolhuis, Phys.\ Rev.\ E {\bf 59},  618  (1999).

\bibitem{BarHan86}
J.~L. Barrat and J.~P. Hansen, J.\ Phys.\ (Paris) {\bf 47},  1547  (1986).

\bibitem{McrHay88}
R. {McRae} and A.~D.~J. Haymet, J.\ Chem.\ Phys. {\bf 88},  1114  (1988).

\bibitem{Pusey87}
P.~N. Pusey, J.\ Phys.\ (Paris) {\bf 48},  709  (1987).

\bibitem{Bartlett97}
P. Bartlett, J.\ Chem.\ Phys. {\bf 107},  188  (1997).

\bibitem{Bartlett98}
P. Bartlett, J.\ Chem.\ Phys. {\bf 109},  10970  (1998).

\bibitem{Sear98}
R.~P. Sear, Europhys.\ Lett. {\bf 44},  531  (1998).

\bibitem{BarWar99}
P. Bartlett and P.~B. Warren, Phys.\ Rev.\ Lett. {\bf 82},  1979  (1999).

\bibitem{XuBau03}
H. Xu and M. Baus, J.\ Chem.\ Phys. {\bf 118},  5045  (2003).

\bibitem{Sollich02}
P. Sollich, J.\ Phys.\ Cond.\ Matt. {\bf 14},  R79  (2002).

\bibitem{Bartlett00}
P. Bartlett, J.\ Phys.\ Cond.\ Matt. {\bf 12},  A275  (2000).

\bibitem{PusVan87}
P.~N. Pusey and W. {van Megen}, Phys.\ Rev.\ Lett. {\bf 59},  2083  (1987).

\bibitem{AueFre01}
S. Auer and D. Frenkel, Nature {\bf 413},  711  (2001).

\bibitem{EvaHol01}
R.~M.~L. Evans and C.~B. Holmes, Phys.\ Rev.\ E {\bf 64},  011404  (2001).

\bibitem{ZhuLiRogMeyOttRusCha97}
J.~X. Zhu {\it et~al.}, Nature {\bf 387},  883  (1997).

\bibitem{FasSol03}
M. Fasolo and P. Sollich, Phys.\ Rev.\ Lett. {\bf 91},  068301  (2003).

\bibitem{SolWarCat01}
P. Sollich, P.~B. Warren, and M.~E. Cates, Adv.\ Chem.\ Phys. {\bf 116},  265
  (2001).

\bibitem{SalSte82}
J.~J. Salacuse and G. Stell, J.\ Chem.\ Phys. {\bf 77},  3714  (1982).

\bibitem{Boublik70}
T. Boublik, J.\ Chem.\ Phys. {\bf 53},  471  (1970).

\bibitem{ManCarStaLel71}
G.~A. Mansoori, N.~F. Carnahan, K.~E. Starling, and T.~W. {Leland, Jr.}, J.\
  Chem.\ Phys. {\bf 54},  1523  (1971).

\bibitem{CarSta69}
N.~F. Carnahan and K.~F. Starling, J.\ Chem.\ Phys. {\bf 51},  2305  (1969).

\bibitem{Bartlett99}
P. Bartlett, Mol.\ Phys. {\bf 97},  685  (1999).

\bibitem{CotParVegMon96}
X. Cottin, E.~P.~A. Paras, C. Vega, and P.~A. Monson, Fluid Phase Equilib. {\bf
  117},  114  (1996).

\bibitem{CotMon95}
X. Cottin and P.~A. Monson, J.\ Chem.\ Phys. {\bf 102},  3354  (1995).

\bibitem{BarBauHan86}
J.~L. Barrat, M. Baus, and J.~P. Hansen, Phys.\ Rev.\ Lett. {\bf 56},  1063
  (1986).

\bibitem{KraFre91}
W.~G.~T. Kranendonk and D. Frenkel, Mol.\ Phys. {\bf 72},  679  (1991).

\bibitem{kirkwood50}
J.~G. Kirkwood, J.\ Chem.\ Phys. {\bf 18},  380  (1950).

\bibitem{ReiFriLeb59}
H. Reiss, H.~L. Frish, and J.~L. Lebowitz, J.\ Chem.\ Phys. {\bf 31},  369
  (1959).

\bibitem{LebHelPar65}
J.~L. Lebowitz, E. Helfand, and E. Paraestagaard, J.\ Chem.\ Phys. {\bf 43},
  774  (1965).

\bibitem{Widom63}
B. Widom, J.\ Chem.\ Phys. {\bf 39},  2808  (1963).

\bibitem{Warren98}
P.~B. Warren, Phys.\ Rev.\ Lett. {\bf 80},  1369  (1998).

\bibitem{SolCat98}
P. Sollich and M.~E. Cates, Phys.\ Rev.\ Lett. {\bf 80},  1365  (1998).

\bibitem{ClaCueSeaSolSpe00}
N. Clarke {\it et~al.}, J.\ Chem.\ Phys. {\bf 113},  5817  (2000).

\bibitem{SpeSol02}
A. Speranza and P. Sollich, J.\ Chem.\ Phys. {\bf 117},  5421  (2002).

\bibitem{SpeSol03a}
A. Speranza and P. Sollich, J.\ Chem.\ Phys. {\bf 118},  5213  (2003).

\bibitem{SpeSol03b}
A. Speranza and P. Sollich, Phys.\ Rev.\ E {\bf 67},  061702  (2003).

\bibitem{Warren99}
P.~B. Warren, Europhys.\ Lett. {\bf 46},  295  (1999).

\bibitem{Cuesta99}
J.~A. Cuesta, Europhys.\ Lett. {\bf 46},  197  (1999).

\bibitem{Bishop95}
C.~M. Bishop, {\em Neural Networks for Pattern Recognition} (Oxford University
  Press, Oxford, 1995).

\bibitem{EvaFaiPoo98}
R.~M.~L. Evans, D.~J. Fairhurst, and W.~C.~K. Poon, Phys.\ Rev.\ Lett. {\bf
  81},  1326  (1998).

\bibitem{Evans01}
R.~M.~L. Evans, J.\ Chem.\ Phys. {\bf 114},  1915  (2001).

\bibitem{Kofke93}
D.~A. Kofke, J.\ Chem.\ Phys. {\bf 98},  4149  (1993).

\bibitem{WilSol02}
N.~B. Wilding and P. Sollich, J.\ Chem.\ Phys. {\bf 116},  7116  (2002).

\bibitem{Sear99c}
R.~P. Sear, Phys.\ Rev.\ Lett. {\bf 82},  4244  (1999).

\end{thebibliography}
\bibliographystyle{prsty}

\end{document}